\documentclass[twocolumn,english,aps,superscriptaddress,floatfix,final,showpacs,prl]{revtex4}
\usepackage{mathptmx}

\usepackage[T1]{fontenc}
\usepackage[latin9]{inputenc}
\usepackage{graphicx}
\usepackage{esint}

\makeatletter

\newcommand{\lyxdot}{.}

\@ifundefined{textcolor}{}
{%
 \definecolor{BLACK}{gray}{0}
 \definecolor{WHITE}{gray}{1}
 \definecolor{RED}{rgb}{1,0,0}
 \definecolor{GREEN}{rgb}{0,1,0}
 \definecolor{BLUE}{rgb}{0,0,1}
 \definecolor{CYAN}{cmyk}{1,0,0,0}
 \definecolor{MAGENTA}{cmyk}{0,1,0,0}
 \definecolor{YELLOW}{cmyk}{0,0,1,0}
}

\renewcommand\[{\begin{equation}}
\renewcommand\]{\end{equation}} 

\makeatother

\usepackage{babel}
\begin{document}

\title{Mottness-induced healing in strongly correlated superconductors}

\author{Shao Tang }

\affiliation{Department of Physics and National High Magnetic Field Laboratory,
Florida State University, Tallahassee, Florida 32306, USA}

\author{E. Miranda}

\affiliation{Instituto de Física Gleb Wataghin, Campinas State University, Rua
Sérgio Buarque de Holanda, 777, CEP 13083-859, Campinas, Brazil}

\author{V. Dobrosavljevic}

\affiliation{Department of Physics and National High Magnetic Field Laboratory,
Florida State University, Tallahassee, Florida 32306, USA}
\begin{abstract}
We study impurity healing effects in models of strongly correlated
superconductors. We show that in general both the range and the amplitude
of the spatial variations caused by nonmagnetic impurities are significantly
suppressed in the superconducting as well as in the normal states.
We explicitly quantify the weights of the local and the non-local
responses to inhomogeneities and show that the former are overwhelmingly
dominant over the latter. We find that the local response is characterized
by a well-defined healing length scale, which is restricted to only
a few lattice spacings over a significant range of dopings in the
vicinity of the Mott insulating state. We demonstrate that this healing
effect is ultimately due to the suppression of charge fluctuations
induced by Mottness. We also define and solve analytically a simplified
yet accurate model of healing, within which we obtain simple expressions
for quantities of direct experimental relevance.
\end{abstract}


\maketitle
\textit{Introduction}.---Strong electronic correlations are believed
to be essential for a complete understanding of many classes of unconventional
superconductors, such as the cuprates \cite{anderson1987resonating,anderson2004physics,Dagotto08072005,lee2006doping},
heavy fermion superconductors \cite{varma85}, organic materials \cite{powellmckenzie06,powellmckenzie11}
and iron pnictides \cite{johnston10}. Among the many puzzling features
of these systems is their behavior in the presence of disorder. In
the case of the cuprates, experiments have shown that these \textit{$d$}\textit{\emph{-wave
superconductors are quite robust against disorder as introduced by
carrier doping \cite{Dagotto08072005,McElroy12082005,fujita2012spectroscopic}.
In particular, there seems to be a ``quantum protection'' of the $d$-wave
nodal points \cite{Anderson21042000}. Other anomalies were found
in the organics \cite{analytisetal06} and the pnictides \cite{lietal12}.
Although it is controversial whether conventional theory is able to
explain these features}} \cite{balatsky2006impurity}\textit{\emph{,}}
strong electronic interactions can give rise to these impurity screening
effects. Indeed, they have been captured numerically by\textit{\emph{
the Gutzwiller-projected wave function \cite{garg2008strong,Fukushima20083046,PhysRevB.79.184510},
even though a deeper insight into the underlying mechanism is still
lacking. Similar impurity screening phenomena have been found as a
result of strong correlations in the metallic state of the Hubbard
model \cite{PhysRevLett.104.236401}.}}

Despite this progress, it would be desirable to understand to what
extent this disorder screening is due only to the presence of strong
correlations or whether it is dependent on the details of the particular
model or system.\textit{\emph{ For example, are the effects of the
inter-site super-exchange, crucial to describe the cuprates, essential
for this phenomenon? To address these issues, it would be fruitful
to have an analytical treatment of the problem. We will describe in
this Letter how an expansion in the disorder potential is able to
provide important insights into these questions. In particular, we
show that the ``healing'' of the impurities is a sheer consequence
of the strong correlations and depend very little on the symmetry
of the superconducting (SC) state or the inclusion of inter-site magnetic
correlations.}}

We considered dilute nonmagnetic impurities in an otherwise homogenous,
strongly correlated electronic state. We avoided complications related
to the nucleation of possible different competing orders by the added
impurities, such as fluctuating or static charge- and spin-density-waves
\cite{fradkin2012high,Ghiringhelli17082012,PhysRevLett.96.017007,ubbens1992flux}
or the formation of local moments \cite{alloul2009defects}. Therefore,
we focused only on how a given strongly correlated state readjusts
itself in the presence of the impurities. We used a spatially inhomogeneous
slave boson treatment \textit{\emph{\cite{coleman1984new,kotliar1986new,lee2006doping,kotliar1988superexchange,lee19982},
which allowed us to perform a complete quantitative calculation. We
have allowed for either or both of $d$-wave SC and $s$-wave resonating
valence bond (RVB) orders.}}

Our analytical and numerical results demonstrate that (i) for sufficiently
weak correlations we recover the results of the conventional theory
\cite{balatsky2006impurity}, in which the variations of the different
fields induced by the impurities show oscillations with a long-ranged
power-law envelope; (ii) for strong interactions and in several different
broken symmetry states, the amplitude of the oscillations is strongly
suppressed by a common pre-factor $x$, the deviation from half-filling;
(iii) the spatial disturbances of the SC gap are healed over a precisely
defined length scale, which does not exceed a few lattice parameters
around the impurities; and (iv) this ``healing effect'' is intrinsically
tied to the proximity to the Mott insulating state, even though it
survives up to around 30\% doping.

\begin{figure*}[t]
\includegraphics[scale=0.7]{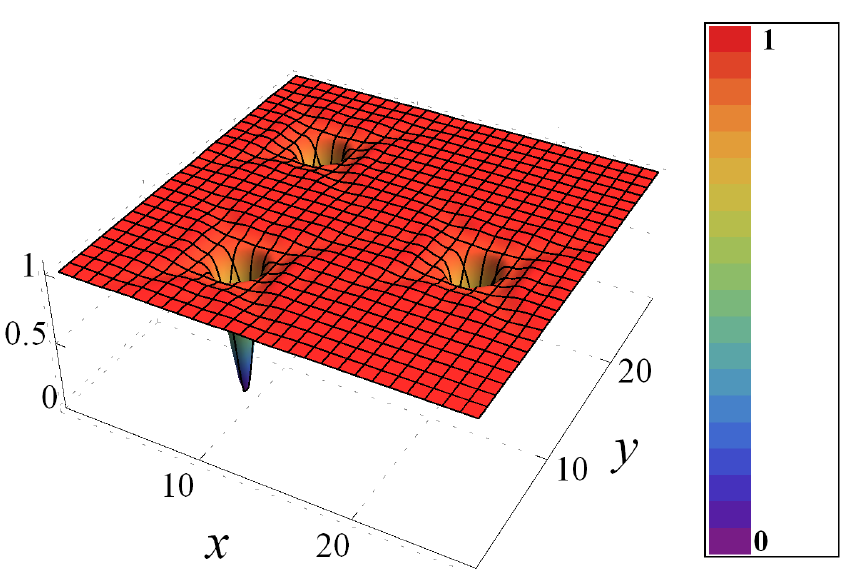}\includegraphics[scale=1.3]{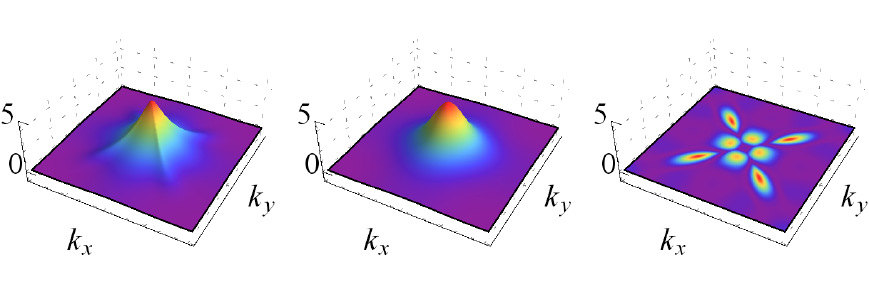}

\includegraphics[scale=0.7]{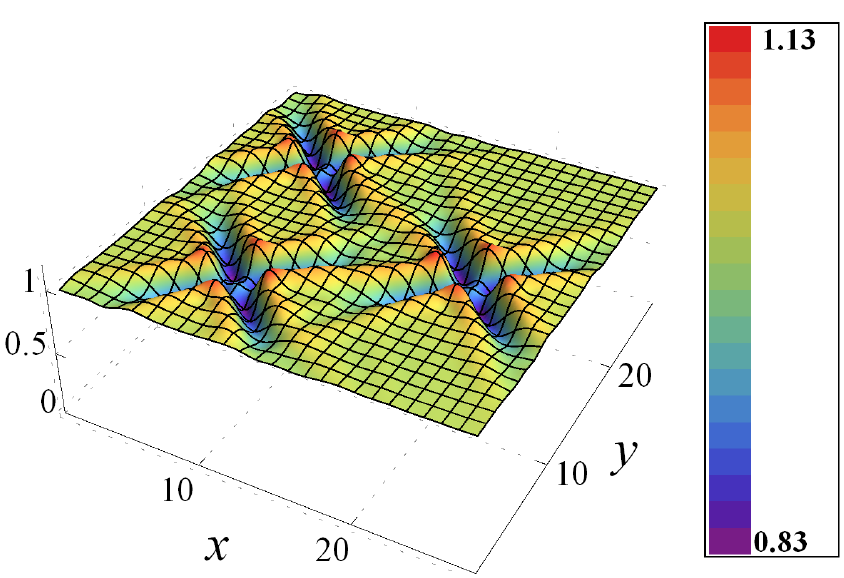}\includegraphics[scale=1.3]{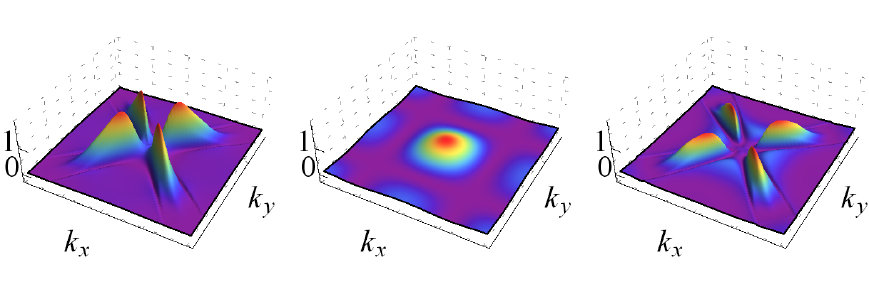}

\protect\caption{\label{fig:gapfluctuations}Spatial variations of normalized local
SC gap $\frac{\Delta_{i}}{\Delta_{0}}$ for three impurities (first
column) and the corresponding power spectra $S(\mathbf{k})$, $S(\mathbf{k})_{loc}$
and $S(\mathbf{k})_{nonloc}$ (second to fourth columns), in the presence
(top) and in the absence (bottom) of correlations for $x=0.2$. The
strong suppression of gap oscillations by correlations can be traced
to the dominance of the local, spherically symmetric power spectrum
{[}$S_{loc}\left(\mathbf{k}\right)${]} over the non-local anisotropic
part {[}$S_{nonloc}\left(\mathbf{k}\right)${]}.}
\end{figure*}

\textit{Model and method}.---We study the $t-t^{\prime}-J$ model
on a cubic lattice in $d$ dimensions with dilute nonmagnetic impurities

\begin{equation}
H=-\sum_{ij\sigma}t_{ij}c_{i\sigma}^{\dagger}c_{j\sigma}+J\sum_{ij}\mathbf{S}_{i}\cdot\mathbf{S}_{j}+\sum_{i}(\epsilon_{i}-\mu_{0})n_{i},
\end{equation}
where $t_{ij}$ are the hopping matrix elements between nearest-neighbor
($t$) and second-nearest-neighbor ($t^{\prime}$) sites, $c_{i\sigma}^{\dagger}\left(c_{i\sigma}\right)$
is the creation (annihilation) operator of an electron with spin projection
$\sigma$ at site $i$, $J$ is the super-exchange coupling constant
between nearest-neighbor sites, $n_{i}=\sum_{\sigma}c_{i\sigma}^{\dagger}c_{i\sigma}$
is the number operator, $\mu_{0}$ is the chemical potential and $\epsilon_{i}$
is the impurity potential. The no double occupancy constraint ($n_{i}\le1$)
is implied. We set the nearest-neighbor hopping $t$ as the energy
unit and choose $t^{\prime}=-0.25t$. To treat this model, we employ
the $U(1)$ slave boson theory \cite{coleman1984new,kotliar1988superexchange,lee2006doping,PhysRevB.36.857}.
Details can be found in \cite{lee2006doping} and we only describe
it very briefly here. It starts with the replacement $c_{i\sigma}^{\dagger}=f_{i\sigma}^{\dagger}b_{i}$,
where $f_{i\sigma}^{\dagger}$ and $b_{i}$ are auxiliary fermionic
(spinon) and bosonic fields, and the representation is faithful in
the subspace $n_{i}\le1$ if the constraint $\sum_{\sigma}f_{i\sigma}^{\dagger}f_{i\sigma}+b_{i}^{\dagger}b_{i}=1$
is enforced. This is implemented by a Lagrange multiplier $\lambda_{i}$
on each site. The $J$ term is then decoupled by Hubbard-Stratonovich
fields in the particle-particle ($\Delta_{ij}$) and particle-hole
($\chi_{ij}$) channels. The auxiliary bosonic fields are all treated
in the saddle-point approximation: $\left\langle b_{i}\right\rangle =r_{i}=\sqrt{Z_{i}}$
gives the quasiparticle residue, $\left\langle \lambda_{i}\right\rangle $
renormalizes the site energies and $\chi_{ij}=\sum_{\sigma}\left\langle f_{i\sigma}^{\dagger}f_{j\sigma}\right\rangle $
and $\Delta_{ij}=\left\langle f_{i\uparrow}f_{j\downarrow}-f_{i\downarrow}f_{j\uparrow}\right\rangle $
describe, respectively, the strength of a spinon singlet and the pairing
amplitude across the corresponding bonds. Note that we do not assume
these values are spatially uniform. This treatment is equivalent to
the Gutzwiller approximation \cite{anderson2004physics,garg2008strong}.
In terms of Gorkov's spinor notation \cite{abrikosov1975methods}
with $\Psi_{i}(i\omega_{n})=\left[\begin{array}{cc}
f_{i\uparrow}^{\dagger}(i\omega_{n}) & f_{i\downarrow}(-i\omega_{n})\end{array}\right]^{\dagger}$, where $\omega_{n}$ is the fermionic Matsubara frequency, the spinon
Green's function is a $2\times2$ matrix: $\left[G_{ij}(i\omega_{n})\right]_{ab}=-\left\langle \Psi_{i}(i\omega_{n})\Psi_{j}^{\dagger}(i\omega_{n})\right\rangle _{ab}$.
Defining $h_{ij}\equiv-t_{ij}$, the saddle-point equations read as
follows

\begin{eqnarray}
\chi_{ij} & = & 2T\sum_{n}(G_{ij})_{11},\label{eq:chieq}\\
\Delta_{ij} & = & -2T\sum_{n}(G_{ij})_{12},\label{eq:order parameter}
\end{eqnarray}

\begin{eqnarray}
(r_{i}^{2}-1) & = & -2T\sum_{n}(G_{ii})_{11},\label{eq:constraint}\\
\lambda_{i}r_{i} & = & -2T\sum_{nl}h_{il}r_{l}(G_{il})_{11}=-\sum_{l}h_{il}r_{l}\chi_{il}.\label{eq:sum rule}
\end{eqnarray}

Note that we used Eq.~(\ref{eq:chieq}) in the second equality of
Eq.~(\ref{eq:sum rule}). At $T=0$ and in the clean limit $\epsilon_{i}=0$,
we have $Z=Z_{0}=x$. The Mott metal-insulator transition is signaled
by the vanishing of the quasi-particle weight $Z_{0}\rightarrow0$
at half-filling. It will be interesting to compare the results of
the above procedure with the ones obtained from solving only Eqs.~(\ref{eq:chieq}-\ref{eq:order parameter})
while setting $Z_{i}=1$ and $\lambda_{i}=0$. The two sets will be
called correlated and non-correlated, respectively. In order to be
able to compare them, we set $J=t/3$ in the correlated case and adjusted
$J$ in the non-correlated case in such a way that the two clean dimensionful
SC gaps coincide, as discussed in reference \cite{garg2008strong}.

\textit{Healing}.--- Although the detailed solutions of Eqs.~(\ref{eq:chieq}-\ref{eq:sum rule})
can be straightforwardly obtained numerically, we will focus on the
case of weak scattering by dilute impurities and expand those equations
up to first order in $\varepsilon_{i}$ around the homogeneous case.
It has been shown and we confirm that disorder induces long-ranged
oscillations in various physical quantities, specially near the nodal
directions in the $d$-wave SC state \cite{balatsky2006impurity}.
The linear approximation we employ is quite accurate for these extended
disturbances far from the impurities, since these are always small.
Besides, it provides more analytical insight into the results.

In general, we can expand the spatial variations of the various order
parameters in different symmetry channels through cubic harmonics:
$\delta\varphi_{ij}=\sum_{g}\delta\varphi_{i}\Gamma(g)_{ij}$ where
$\varphi_{ij}=\chi_{ij}$ or $\Delta_{ij}$ and $\Gamma(g)_{ij}$
are the basis functions for cubic harmonic $g$ of the square lattice
\footnote{$s$, $d_{x^{2}-y^{2}}$, $d_{xy}$, etc., with basis functions expressed
as: $\cos k_{x}+\cos k_{y}$, $\cos k_{x}-\cos k_{y}$ and $\sin k_{x}\sin k_{y}$,
etc.%
}. In the current discussion, we choose $\delta\chi_{ij}=\delta\chi_{i}\Gamma(s)_{ij}$
and $\delta\Delta_{ij}=\delta\Delta_{i}\Gamma(d_{x^{2}-y^{2}})_{ij}$,
as we are interested in oscillations with the same symmetry as the
ground state \cite{coleman1984new,kotliar1988superexchange,lee2006doping,PhysRevB.36.857}.
We also assume there is no phase difference between order parameters
on different bonds linked to same site. Then, we can define ``local''
spatial variations of the order parameters as $\delta\chi_{i}\equiv\frac{1}{2d}\sum_{j}\delta\chi_{ij}\Gamma(s)_{ij}$
and $\delta\Delta_{i}\equiv\frac{1}{2d}\sum_{j}\delta\Delta_{ij}\Gamma(d_{x^{2}-y^{2}})_{ij}$.
Details of the calculation can be found in the Supplemental Material
\cite{Suppl}.

We find that both $\delta\chi_{ij}$ and $\delta\Delta_{ij}$, as
well as the impurity-induced charge disturbance $\delta n_{i}$, are
proportional to $Z_{0}=x$, indicating the importance of strong correlations
for the healing effect. Indeed, we can trace back this behavior to
the readjustment of the $r_{i}$ and $\lambda_{i}$ fields, as encoded
in Eqs.~(\ref{eq:constraint}-\ref{eq:sum rule}). Besides, this
${\cal O}\left(x\right)$ suppression is a generic consequence of
the structure of the mean-field equations and holds for different
broken symmetry states, such as the flux phase state, $s$-wave superconductivity,
etc.

Let us focus in more detail on the spatial variations of the local
pairing field $\delta\Delta_{i}$. In the first column of Fig.~\ref{fig:gapfluctuations}
we show results for $\delta\Delta_{i}$ for three identical impurities.
The ``cross-like'' tails near the nodal directions \cite{PhysRevLett.76.2386}
are conspicuous in the absence of correlations (bottom) but are strongly
suppressed in their presence (top). While this suppression is further
enhanced as the Mott metal-insulator transition is approached ($x\to0$),
it is still quite significant even at optimal doping ($x=0.2$). This
is the ``healing'' effect previously reported \textit{\emph{\cite{garg2008strong,Fukushima20083046,PhysRevB.79.184510}.}}
In order to gain insight into its underlying mechanism, we look at
the spatial correlation function of local gap fluctuations 
\begin{equation}
\left\langle \frac{\delta\Delta_{i}}{\Delta_{0}}\frac{\delta\Delta_{j}}{\Delta_{0}}\right\rangle _{disorder}=f\left(\mathbf{r}_{i}-\mathbf{r}_{j}\right),\label{eq:gapcorrfunc}
\end{equation}
where the brackets denote an average over disorder, after which lattice
translation invariance is recovered. The Fourier transform of $f\left(\mathbf{r}\right)$
can be written in the linear approximation as\vspace{-18pt}

\begin{equation}
f\left(\mathbf{k}\right)=\alpha W^{2}S\left(\mathbf{k}\right),\label{eq:gapcorrfuncink}
\end{equation}
where $W$ is the disorder strength, $\alpha$ depends on the detailed
bare disorder distribution, and the ``power spectrum'' (PS) $S\left(\mathbf{k}\right)$
is related to gap linear response function $M_{\Delta}\left(\mathbf{k}\right)$
by $S\left(\mathbf{k}\right)=M_{\Delta}^{2}\left(\mathbf{k}\right)$.
The latter is defined by Fourier transforming the kernel in $\delta\Delta_{i}=\Delta_{0}\sum_{j}M_{\Delta}\left(\mathbf{r}_{i}-\mathbf{r}_{j}\right)\varepsilon_{j}$,
which in turn can be easily obtained from the solution of the linearized
equations \cite{Suppl}. Inspired by the strongly localized gap fluctuations
at the top left of Fig.~\ref{fig:gapfluctuations}, we define the
local component of the PS $S_{loc}\left(\mathbf{k}\right)\equiv M_{\Delta,loc}^{2}\left(\mathbf{k}\right)$,
where $M_{\Delta,loc}\left(\mathbf{k}\right)$ is obtained by \emph{restricting
the lattice sums up to the second nearest neighbor distance} ($\sqrt{2}a$)
in the linearized equations \cite{Suppl}. We also define $S_{nonloc}\left(\mathbf{k}\right)=M_{\Delta,nonloc}^{2}\left(\mathbf{k}\right)\equiv\left[M_{\Delta}\left(\mathbf{k}\right)-M_{\Delta,loc}\left(\mathbf{k}\right)\right]^{2}$.
In the last three columns of Fig.~\ref{fig:gapfluctuations}, we
show, in this order, $S\left(\mathbf{k}\right)$, $S_{loc}\left(\mathbf{k}\right)$,
and $S_{nonloc}\left(\mathbf{k}\right)$ for the correlated (top)
and non-correlated (bottom) cases at $x=0.2$. Clearly, in the presence
of correlations the local PS is characterized by a smooth, spherically
symmetric bell-shaped function, whereas the non-local part is highly
anisotropic. Besides and more importantly, the non-local PS is negligibly
small in the correlated case. The full PS is thus \emph{overwhelmingly
dominated} by the local part, unlike in the non-correlated case. In
the Supplemental Material \cite{Suppl}, we extend the analysis to
the underdoped and overdoped regimes, where very similar behavior
is found, even up to dopings of $x=0.3$. 

\begin{figure}
\includegraphics[scale=0.8]{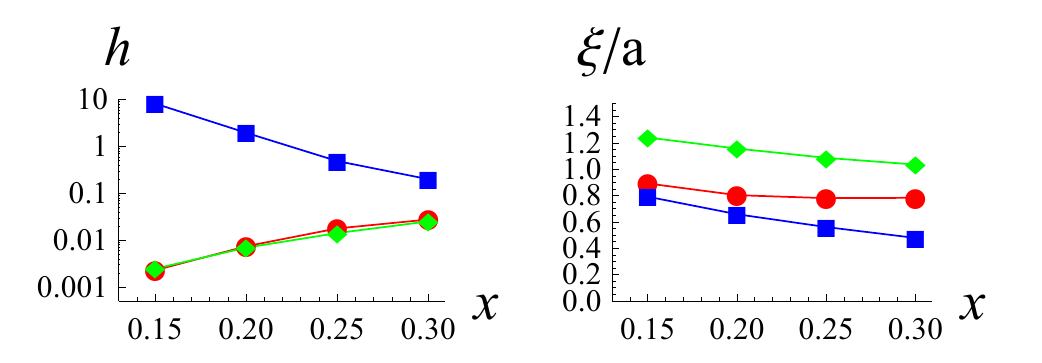}

\protect\caption{\label{fig:healcsiandh}Left panel: the healing factor $h$ as a function
of doping in the uncorrelated case (blue curve with squares), in the
correlated case (red curve with circles), and in the correlated case
without $\delta\chi_{i}$ fluctuations (green curve with diamonds).
Right panel: doping dependence of the SC ($\xi_{S}$, red curve with
circles) and normal state ($\xi_{N}$, blue curve with squares) healing
lengths. The green curve with diamonds gives $\xi_{S}$ calculated
within the minimal model (see text).}
\end{figure}

In order to quantify the localized nature of the healing effect, we
are led to a natural definition of a ``healing factor'' $h$ in
the $d$-wave SC state\vspace{-12pt}

\begin{equation}
h=\frac{\int S_{nonloc}\left(\mathbf{k}\right)d^{2}k}{\int S_{loc}\left(\mathbf{k}\right)d^{2}k},
\end{equation}
where the integration is over the first Brillouin zone. It measures
the relative weight of non-local and local parts of the gap PS. The
healing factor as a function of doping is shown on the left panel
of Fig.~\ref{fig:healcsiandh} for the non-correlated (blue) and
correlated (red) cases. The contrast is striking. When correlations
are present, $h$ is extremely small up to 30\% doping and the gap
disturbance is restricted to a small area around the impurities. In
contrast, without correlations significant pair fluctuations occur
over quite a large area for all dopings shown. We conclude that the
strong dominance of the local part over the highly anisotropic non-local
contribution caused by correlations is the \emph{key feature behind
the healing process}.

The shape of $S_{loc}\left(\mathbf{k}\right)$ shows that the gap
disturbance created by an impurity is healed over a well-defined distance,
the ``healing length'' $\xi_{S}$. This length scale can be obtained
by expanding the inverse of $M_{\Delta,loc}\left(\mathbf{k}\right)$
{[}or equivalently $M_{\Delta}\left(\mathbf{k}\right)${]} up to second
order in $k^{2}$, thus defining a Lorentzian in $\mathbf{k}$-space
\begin{equation}
M_{\Delta,loc}\left(\mathbf{k}\right)\approx\frac{1}{A+Bk^{2}}.\label{eq:lorentzian}
\end{equation}
The SC healing length is then given by $\xi_{S}=\sqrt{B/A}$. The
$x$ dependence of $\xi_{S}$ is shown in red on the right panel of
Fig.~\ref{fig:healcsiandh}. It is of the order of one lattice spacing
in the relevant range $0.15<x<0.3$. It should be noted that precisely
the same length scale also governs the healing of charge fluctuations
in the SC state, showing that this phenomenon is generic to the strongly
correlated state. A similar procedure can be carried out for the charge
fluctuations in the normal state, thus defining a normal state healing
length $\xi_{N}$ \cite{Suppl}. The blue curve of the right panel
of Fig.~\ref{fig:healcsiandh} shows the $x$ dependence of $\xi_{N}$,
which is also of the order of one lattice spacing. 

\textit{Mottness-induced healing.---}The healing effect we have described
comes almost exclusively from the $\delta r_{i}$ and $\delta\lambda_{i}$
fluctuations: $h$ is hardly affected by the $\delta\chi_{i}$ field.
If we suppress the $\delta\chi_{i}$ fluctuations completely \cite{Suppl},
there is only a tiny change in the results, as shown by the green
curve of the left panel of Fig~\ref{fig:healcsiandh}. The same is
not true, however, if we turn off either $\delta r_{i}$ or $\delta\lambda_{i}$
or both. We conclude that the healing effect in the $d$-wave SC state
originates from the strong correlation effects alone, rather than
the spinon correlations.

\begin{figure}
\includegraphics{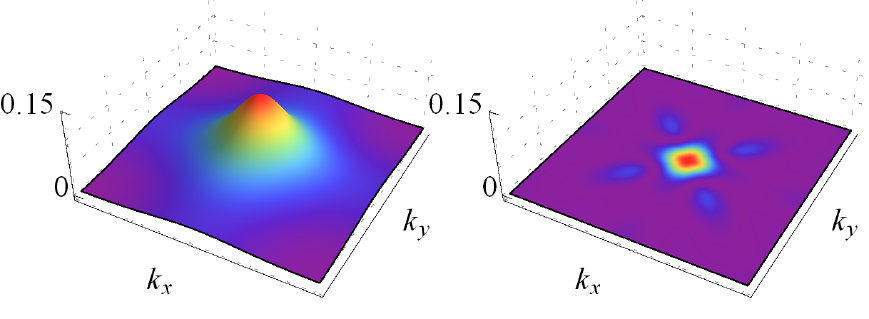}

\protect\caption{\label{fig:densityfluct}Local (left) and nonlocal (right) parts of
the charge-fluctuation power spectra $N(\mathbf{k})_{loc}$ and $N(\mathbf{k})_{nonloc}$
in the presence of strong correlations for $x=0.2$.}
\end{figure}

Within the linear approximation we are employing, all fluctuation
fields ($\delta\Delta$, $\delta r$, etc.) are proportional, in $\mathbf{k}$-space,
to the disorder potential $\varepsilon\left(\mathbf{k}\right)$. Therefore,
they are also proportional to each other. In particular, given the
centrality of the strong correlation fields, it is instructive to
write the gap fluctuations in terms of the slave boson fluctuations
\begin{equation}
\delta\Delta\left(\mathbf{k}\right)=-2\chi_{pc}\left(\mathbf{k}\right)r\delta r\left(\mathbf{k}\right)=\chi_{pc}\left(\mathbf{k}\right)\delta n\left(\mathbf{k}\right).\label{eq:MMgapfluc}
\end{equation}
In the last equality, we used $n_{i}=1-r_{i}^{2}$, which enables
us to relate two physically transparent quantities: the gap and the
charge fluctuations. Indeed, this will provide crucial physical insight
into the healing process. By focusing on the linear charge response
to the disorder potential $\delta n\left(\mathbf{k}\right)=n_{0}M_{n}\left(\mathbf{k}\right)\varepsilon\left(\mathbf{k}\right)$,
we can, in complete analogy with the gap fluctuations, define a PS
for the spatial charge fluctuations, $N\left(\mathbf{k}\right)=M_{n}^{2}\left(\mathbf{k}\right)$.
This PS can also be broken up into local {[}$N_{loc}\left(\mathbf{k}\right)=M_{n,loc}^{2}\left(\mathbf{k}\right)${]}
and non-local \{$N_{nonloc}\left(\mathbf{k}\right)=\left[M_{n}\left(\mathbf{k}\right)-M_{n,loc}\left(\mathbf{k}\right)\right]^{2}$\}
parts, as was done for the gap-fluctuation PS. These two contributions,
obtained from the solution of the full linearized equations, are shown
in Fig.~\ref{fig:densityfluct}. The charge PS in the correlated
$d$-wave SC state is also characterized by a smooth, almost spherically
symmetric local part and a negligibly small anisotropic non-local
contribution. Note also the strong similarity between the local PS
for gap (top row of Fig.~\ref{fig:gapfluctuations}) and charge fluctuations.
This shows a strong connection between the gap and charge responses.
Evidently, this is also reflected in real space, where the charge
disturbance is healed in the same strongly localized fashion as the
gap disturbance \cite{Suppl}. In fact, the local part of the charge
response function $M_{n,loc}\left(\mathbf{k}\right)$ can be shown
to be well approximated by a Lorentzian \cite{Suppl} and we can write
for small $\mathbf{k}$
\begin{equation}
\delta\Delta_{loc}\left(\mathbf{k}\right)\approx-\chi_{pc}\left(\mathbf{k}=0\right)\frac{8r^{2}/\lambda}{k^{2}+\xi_{S}^{-2}}\varepsilon\left(\mathbf{k}\right),\label{eq:MMgapflucloc}
\end{equation}
where the SC healing length $\xi_{S}$ can be expressed in terms of
the Green's functions of the clean system \cite{Suppl}. The relations
implied by Eqs.~(\ref{eq:MMgapfluc}) and (\ref{eq:MMgapflucloc}),
as well as the doping dependence of the quantities in them, could
be tested in STM studies and would constitute an important test of
this theory. 

Eqs.~(\ref{eq:MMgapfluc}-\ref{eq:MMgapflucloc}) allow us to obtain
a clear physical picture of the healing mechanism. The spatial gap
fluctuations can be viewed as being ultimately determined by the charge
fluctuations. Furthermore, their ratio $\chi_{pc}\left(\mathbf{k}\right)$,
which is essentially a pair-charge correlation function, is a rather
smooth function \emph{of order unity},\emph{ only weakly renormalized
by interactions}. Therefore, it is the strong suppression of charge
fluctuations by ``Mottness'', as signaled by the $r^{2}$ factor in
Eq.~(\ref{eq:MMgapflucloc}), which is behind the healing of gap
fluctuations. This elucidates the physics of healing previously found
numerically \textit{\emph{\cite{garg2008strong,Fukushima20083046,PhysRevB.79.184510}.
}}It also suggests that the healing phenomenon is generic to Mott
systems \cite{PhysRevLett.104.236401} and is not tied to the specifics
of the cuprates.

\textit{A minimal model.---}Interestingly, the crucial role played
by the strong correlation fields ($r_{i}$ and $\lambda_{i}$) suggests
a ``minimal model'' (MM) for an accurate description of the healing
process, which we define as follows: (i) the spatially fluctuating
strong correlation fields $r_{i}$ and $\lambda_{i}$ are first calculated
for the self-consistently determined, \emph{fixed, uniform} $\Delta$
and $\chi$, and then (ii) the effects of their spatial readjustments
are fed back into the gap equation~(\ref{eq:order parameter}) in
order to find $\delta\Delta_{i}$ \cite{Suppl}. The accuracy of this
procedure can be ascertained by the behavior of the healing factor:
it is \emph{numerically indistinguishable} from the green curve of
the left panel of Fig.~\ref{fig:healcsiandh}. Furthermore, the value
of $\xi_{S}$ calculated within the MM differs from the one obtained
from the solution of the full linearized equations by at most 20\%
(red and green curves on the right panel of Fig.~\ref{fig:healcsiandh}).
Besides its accuracy, the advantage of this MM description lies in
the simplicity of the analytical expressions obtained. As shown in
the Supplemental Material \cite{Suppl}, it provides simple expressions
for the important quantities $\chi_{pc}\left(\mathbf{k}\right)$ and
$\xi_{S}$.

\textit{Conclusions.---}In this work, we have found an inextricable
link between the healing of gap and charge disturbances in strongly
correlated superconductors, suggesting that this phenomenon is generic
to any system close to Mott localization. An important experimental
test of this link would be provided by STM studies of the organic
superconductors \textit{\emph{\cite{analytisetal06} and maybe the
pnictides \cite{lietal12}. Whether it is also relevant for heavy
fermion systems \cite{FigginsMorr2011} is an open question left for
future study. }}

We acknowledge support by CNPq through grant 304311/2010-3 (EM), FAPESP
through grant 07/57630- 5 (EM) and NSF through grant DMR-1005751 (ST
and VD).

\bibliographystyle{apsrev}
\bibliography{manuscript}

\begin{thebibliography}{32}
\expandafter\ifx\csname natexlab\endcsname\relax\def\natexlab#1{#1}\fi
\expandafter\ifx\csname bibnamefont\endcsname\relax
  \def\bibnamefont#1{#1}\fi
\expandafter\ifx\csname bibfnamefont\endcsname\relax
  \def\bibfnamefont#1{#1}\fi
\expandafter\ifx\csname citenamefont\endcsname\relax
  \def\citenamefont#1{#1}\fi
\expandafter\ifx\csname url\endcsname\relax
  \def\url#1{\texttt{#1}}\fi
\expandafter\ifx\csname urlprefix\endcsname\relax\def\urlprefix{URL }\fi
\providecommand{\bibinfo}[2]{#2}
\providecommand{\eprint}[2][]{\url{#2}}

\bibitem[{\citenamefont{Anderson}(1987)}]{anderson1987resonating}
\bibinfo{author}{\bibfnamefont{P.}~\bibnamefont{Anderson}},
  \bibinfo{journal}{Science} \textbf{\bibinfo{volume}{235}},
  \bibinfo{pages}{1196} (\bibinfo{year}{1987}).

\bibitem[{\citenamefont{Anderson et~al.}(2004)\citenamefont{Anderson, Lee,
  Randeria, Rice, Trivedi, and Zhang}}]{anderson2004physics}
\bibinfo{author}{\bibfnamefont{P.}~\bibnamefont{Anderson}},
  \bibinfo{author}{\bibfnamefont{P.}~\bibnamefont{Lee}},
  \bibinfo{author}{\bibfnamefont{M.}~\bibnamefont{Randeria}},
  \bibinfo{author}{\bibfnamefont{T.}~\bibnamefont{Rice}},
  \bibinfo{author}{\bibfnamefont{N.}~\bibnamefont{Trivedi}}, \bibnamefont{and}
  \bibinfo{author}{\bibfnamefont{F.}~\bibnamefont{Zhang}}, \bibinfo{journal}{J.
  Phys.: Condens. Matter} \textbf{\bibinfo{volume}{16}}, \bibinfo{pages}{R755}
  (\bibinfo{year}{2004}).

\bibitem[{\citenamefont{Dagotto}(2005)}]{Dagotto08072005}
\bibinfo{author}{\bibfnamefont{E.}~\bibnamefont{Dagotto}},
  \bibinfo{journal}{Science} \textbf{\bibinfo{volume}{309}},
  \bibinfo{pages}{257} (\bibinfo{year}{2005}).

\bibitem[{\citenamefont{Lee et~al.}(2006)\citenamefont{Lee, Nagaosa, and
  Wen}}]{lee2006doping}
\bibinfo{author}{\bibfnamefont{P.}~\bibnamefont{Lee}},
  \bibinfo{author}{\bibfnamefont{N.}~\bibnamefont{Nagaosa}}, \bibnamefont{and}
  \bibinfo{author}{\bibfnamefont{X.}~\bibnamefont{Wen}}, \bibinfo{journal}{Rev.
  Mod. Phys.} \textbf{\bibinfo{volume}{78}}, \bibinfo{pages}{17}
  (\bibinfo{year}{2006}).

\bibitem[{\citenamefont{Varma}(1985)}]{varma85}
\bibinfo{author}{\bibfnamefont{C.~M.} \bibnamefont{Varma}},
  \bibinfo{journal}{Comments Solid State Phys.} \textbf{\bibinfo{volume}{11}},
  \bibinfo{pages}{221} (\bibinfo{year}{1985}).

\bibitem[{\citenamefont{Powell and McKenzie}(2006)}]{powellmckenzie06}
\bibinfo{author}{\bibfnamefont{B.~J.} \bibnamefont{Powell}} \bibnamefont{and}
  \bibinfo{author}{\bibfnamefont{R.~H.} \bibnamefont{McKenzie}},
  \bibinfo{journal}{J. Phys.: Condens. Matter} \textbf{\bibinfo{volume}{18}},
  \bibinfo{pages}{R827} (\bibinfo{year}{2006}).

\bibitem[{\citenamefont{Powell and McKenzie}(2011)}]{powellmckenzie11}
\bibinfo{author}{\bibfnamefont{B.~J.} \bibnamefont{Powell}} \bibnamefont{and}
  \bibinfo{author}{\bibfnamefont{R.~H.} \bibnamefont{McKenzie}},
  \bibinfo{journal}{Rep. Prog. Phys.} \textbf{\bibinfo{volume}{74}},
  \bibinfo{pages}{056501} (\bibinfo{year}{2011}).

\bibitem[{\citenamefont{Johnston}(2010)}]{johnston10}
\bibinfo{author}{\bibfnamefont{D.~C.} \bibnamefont{Johnston}},
  \bibinfo{journal}{Adv. Phys.} \textbf{\bibinfo{volume}{59}},
  \bibinfo{pages}{803} (\bibinfo{year}{2010}).

\bibitem[{\citenamefont{McElroy et~al.}(2005)\citenamefont{McElroy, Lee,
  Slezak, Lee, Eisaki, Uchida, and Davis}}]{McElroy12082005}
\bibinfo{author}{\bibfnamefont{K.}~\bibnamefont{McElroy}},
  \bibinfo{author}{\bibfnamefont{J.}~\bibnamefont{Lee}},
  \bibinfo{author}{\bibfnamefont{J.~A.} \bibnamefont{Slezak}},
  \bibinfo{author}{\bibfnamefont{D.-H.} \bibnamefont{Lee}},
  \bibinfo{author}{\bibfnamefont{H.}~\bibnamefont{Eisaki}},
  \bibinfo{author}{\bibfnamefont{S.}~\bibnamefont{Uchida}}, \bibnamefont{and}
  \bibinfo{author}{\bibfnamefont{J.~C.} \bibnamefont{Davis}},
  \bibinfo{journal}{Science} \textbf{\bibinfo{volume}{309}},
  \bibinfo{pages}{1048} (\bibinfo{year}{2005}).

\bibitem[{\citenamefont{Fujita et~al.}(2012)\citenamefont{Fujita, Schmidt, Kim,
  Lawler, Lee, Davis, Eisaki, and Uchida}}]{fujita2012spectroscopic}
\bibinfo{author}{\bibfnamefont{K.}~\bibnamefont{Fujita}},
  \bibinfo{author}{\bibfnamefont{A.~R.} \bibnamefont{Schmidt}},
  \bibinfo{author}{\bibfnamefont{E.-A.} \bibnamefont{Kim}},
  \bibinfo{author}{\bibfnamefont{M.~J.} \bibnamefont{Lawler}},
  \bibinfo{author}{\bibfnamefont{D.~H.} \bibnamefont{Lee}},
  \bibinfo{author}{\bibfnamefont{J.}~\bibnamefont{Davis}},
  \bibinfo{author}{\bibfnamefont{H.}~\bibnamefont{Eisaki}}, \bibnamefont{and}
  \bibinfo{author}{\bibfnamefont{S.-i.} \bibnamefont{Uchida}},
  \bibinfo{journal}{J. Phys. Soc. Jap.} \textbf{\bibinfo{volume}{81}},
  \bibinfo{pages}{1005} (\bibinfo{year}{2012}).

\bibitem[{\citenamefont{Anderson}(2000)}]{Anderson21042000}
\bibinfo{author}{\bibfnamefont{P.~W.} \bibnamefont{Anderson}},
  \bibinfo{journal}{Science} \textbf{\bibinfo{volume}{288}},
  \bibinfo{pages}{480} (\bibinfo{year}{2000}).

\bibitem[{\citenamefont{Analytis et~al.}(2006)\citenamefont{Analytis, Ardavan,
  Blundell, Owen, Garman, Jeynes, and Powell}}]{analytisetal06}
\bibinfo{author}{\bibfnamefont{J.~G.} \bibnamefont{Analytis}},
  \bibinfo{author}{\bibfnamefont{A.}~\bibnamefont{Ardavan}},
  \bibinfo{author}{\bibfnamefont{S.~J.} \bibnamefont{Blundell}},
  \bibinfo{author}{\bibfnamefont{R.~L.} \bibnamefont{Owen}},
  \bibinfo{author}{\bibfnamefont{E.~F.} \bibnamefont{Garman}},
  \bibinfo{author}{\bibfnamefont{C.}~\bibnamefont{Jeynes}}, \bibnamefont{and}
  \bibinfo{author}{\bibfnamefont{B.~J.} \bibnamefont{Powell}},
  \bibinfo{journal}{Phys. Rev. Lett.} \textbf{\bibinfo{volume}{96}},
  \bibinfo{pages}{177002} (\bibinfo{year}{2006}).

\bibitem[{\citenamefont{Li et~al.}(2012)\citenamefont{Li, Guo, Zhang, Yuan,
  Tsujimoto, Wang, Sathish, Sun, Yu, Yi et~al.}}]{lietal12}
\bibinfo{author}{\bibfnamefont{J.}~\bibnamefont{Li}},
  \bibinfo{author}{\bibfnamefont{Y.~F.} \bibnamefont{Guo}},
  \bibinfo{author}{\bibfnamefont{S.~B.} \bibnamefont{Zhang}},
  \bibinfo{author}{\bibfnamefont{J.}~\bibnamefont{Yuan}},
  \bibinfo{author}{\bibfnamefont{Y.}~\bibnamefont{Tsujimoto}},
  \bibinfo{author}{\bibfnamefont{X.}~\bibnamefont{Wang}},
  \bibinfo{author}{\bibfnamefont{C.~I.} \bibnamefont{Sathish}},
  \bibinfo{author}{\bibfnamefont{Y.}~\bibnamefont{Sun}},
  \bibinfo{author}{\bibfnamefont{S.}~\bibnamefont{Yu}},
  \bibinfo{author}{\bibfnamefont{W.}~\bibnamefont{Yi}}, \bibnamefont{et~al.},
  \bibinfo{journal}{Phys. Rev. B} \textbf{\bibinfo{volume}{85}},
  \bibinfo{pages}{214509} (\bibinfo{year}{2012}).

\bibitem[{\citenamefont{Balatsky et~al.}(2006)\citenamefont{Balatsky, Vekhter,
  and Zhu}}]{balatsky2006impurity}
\bibinfo{author}{\bibfnamefont{A.}~\bibnamefont{Balatsky}},
  \bibinfo{author}{\bibfnamefont{I.}~\bibnamefont{Vekhter}}, \bibnamefont{and}
  \bibinfo{author}{\bibfnamefont{J.}~\bibnamefont{Zhu}}, \bibinfo{journal}{Rev.
  Mod. Phys.} \textbf{\bibinfo{volume}{78}}, \bibinfo{pages}{373}
  (\bibinfo{year}{2006}).

\bibitem[{\citenamefont{Garg et~al.}(2008)\citenamefont{Garg, Randeria, and
  Trivedi}}]{garg2008strong}
\bibinfo{author}{\bibfnamefont{A.}~\bibnamefont{Garg}},
  \bibinfo{author}{\bibfnamefont{M.}~\bibnamefont{Randeria}}, \bibnamefont{and}
  \bibinfo{author}{\bibfnamefont{N.}~\bibnamefont{Trivedi}},
  \bibinfo{journal}{Nature Phys.} \textbf{\bibinfo{volume}{4}},
  \bibinfo{pages}{762} (\bibinfo{year}{2008}).

\bibitem[{\citenamefont{Fukushima et~al.}(2008)\citenamefont{Fukushima, Chou,
  and Lee}}]{Fukushima20083046}
\bibinfo{author}{\bibfnamefont{N.}~\bibnamefont{Fukushima}},
  \bibinfo{author}{\bibfnamefont{C.-P.} \bibnamefont{Chou}}, \bibnamefont{and}
  \bibinfo{author}{\bibfnamefont{T.~K.} \bibnamefont{Lee}},
  \bibinfo{journal}{J. Phys. Chem. Solids} \textbf{\bibinfo{volume}{69}},
  \bibinfo{pages}{3046 } (\bibinfo{year}{2008}).

\bibitem[{\citenamefont{Fukushima et~al.}(2009)\citenamefont{Fukushima, Chou,
  and Lee}}]{PhysRevB.79.184510}
\bibinfo{author}{\bibfnamefont{N.}~\bibnamefont{Fukushima}},
  \bibinfo{author}{\bibfnamefont{C.-P.} \bibnamefont{Chou}}, \bibnamefont{and}
  \bibinfo{author}{\bibfnamefont{T.~K.} \bibnamefont{Lee}},
  \bibinfo{journal}{Phys. Rev. B} \textbf{\bibinfo{volume}{79}},
  \bibinfo{pages}{184510} (\bibinfo{year}{2009}).

\bibitem[{\citenamefont{Andrade et~al.}(2010)\citenamefont{Andrade, Miranda,
  and Dobrosavljevi\ifmmode~\acute{c}\else
  \'{c}\fi{}}}]{PhysRevLett.104.236401}
\bibinfo{author}{\bibfnamefont{E.~C.} \bibnamefont{Andrade}},
  \bibinfo{author}{\bibfnamefont{E.}~\bibnamefont{Miranda}}, \bibnamefont{and}
  \bibinfo{author}{\bibfnamefont{V.}~\bibnamefont{Dobrosavljevi\ifmmode~\acute{c}\else
  \'{c}\fi{}}}, \bibinfo{journal}{Phys. Rev. Lett.}
  \textbf{\bibinfo{volume}{104}}, \bibinfo{pages}{236401}
  (\bibinfo{year}{2010}).

\bibitem[{\citenamefont{Fradkin and Kivelson}(2012)}]{fradkin2012high}
\bibinfo{author}{\bibfnamefont{E.}~\bibnamefont{Fradkin}} \bibnamefont{and}
  \bibinfo{author}{\bibfnamefont{S.~A.} \bibnamefont{Kivelson}},
  \bibinfo{journal}{Nature Phys.} \textbf{\bibinfo{volume}{8}},
  \bibinfo{pages}{864} (\bibinfo{year}{2012}).

\bibitem[{\citenamefont{Ghiringhelli et~al.}(2012)\citenamefont{Ghiringhelli,
  Le~Tacon, Minola, Blanco-Canosa, Mazzoli, Brookes, De~Luca, Frano, Hawthorn,
  He et~al.}}]{Ghiringhelli17082012}
\bibinfo{author}{\bibfnamefont{G.}~\bibnamefont{Ghiringhelli}},
  \bibinfo{author}{\bibfnamefont{M.}~\bibnamefont{Le~Tacon}},
  \bibinfo{author}{\bibfnamefont{M.}~\bibnamefont{Minola}},
  \bibinfo{author}{\bibfnamefont{S.}~\bibnamefont{Blanco-Canosa}},
  \bibinfo{author}{\bibfnamefont{C.}~\bibnamefont{Mazzoli}},
  \bibinfo{author}{\bibfnamefont{N.~B.} \bibnamefont{Brookes}},
  \bibinfo{author}{\bibfnamefont{G.~M.} \bibnamefont{De~Luca}},
  \bibinfo{author}{\bibfnamefont{A.}~\bibnamefont{Frano}},
  \bibinfo{author}{\bibfnamefont{D.~G.} \bibnamefont{Hawthorn}},
  \bibinfo{author}{\bibfnamefont{F.}~\bibnamefont{He}}, \bibnamefont{et~al.},
  \bibinfo{journal}{Science} \textbf{\bibinfo{volume}{337}},
  \bibinfo{pages}{821} (\bibinfo{year}{2012}).

\bibitem[{\citenamefont{Fang et~al.}(2006)\citenamefont{Fang, Capriotti,
  Scalapino, Kivelson, Kaneko, Greven, and Kapitulnik}}]{PhysRevLett.96.017007}
\bibinfo{author}{\bibfnamefont{A.~C.} \bibnamefont{Fang}},
  \bibinfo{author}{\bibfnamefont{L.}~\bibnamefont{Capriotti}},
  \bibinfo{author}{\bibfnamefont{D.~J.} \bibnamefont{Scalapino}},
  \bibinfo{author}{\bibfnamefont{S.~A.} \bibnamefont{Kivelson}},
  \bibinfo{author}{\bibfnamefont{N.}~\bibnamefont{Kaneko}},
  \bibinfo{author}{\bibfnamefont{M.}~\bibnamefont{Greven}}, \bibnamefont{and}
  \bibinfo{author}{\bibfnamefont{A.}~\bibnamefont{Kapitulnik}},
  \bibinfo{journal}{Phys. Rev. Lett.} \textbf{\bibinfo{volume}{96}},
  \bibinfo{pages}{017007} (\bibinfo{year}{2006}).

\bibitem[{\citenamefont{Ubbens and Lee}(1992)}]{ubbens1992flux}
\bibinfo{author}{\bibfnamefont{M.}~\bibnamefont{Ubbens}} \bibnamefont{and}
  \bibinfo{author}{\bibfnamefont{P.}~\bibnamefont{Lee}},
  \bibinfo{journal}{Phys. Rev. B} \textbf{\bibinfo{volume}{46}},
  \bibinfo{pages}{8434} (\bibinfo{year}{1992}).

\bibitem[{\citenamefont{Alloul et~al.}(2009)\citenamefont{Alloul, Bobroff,
  Gabay, and Hirschfeld}}]{alloul2009defects}
\bibinfo{author}{\bibfnamefont{H.}~\bibnamefont{Alloul}},
  \bibinfo{author}{\bibfnamefont{J.}~\bibnamefont{Bobroff}},
  \bibinfo{author}{\bibfnamefont{M.}~\bibnamefont{Gabay}}, \bibnamefont{and}
  \bibinfo{author}{\bibfnamefont{P.}~\bibnamefont{Hirschfeld}},
  \bibinfo{journal}{Rev. Mod. Phys.} \textbf{\bibinfo{volume}{81}},
  \bibinfo{pages}{45} (\bibinfo{year}{2009}).

\bibitem[{\citenamefont{Coleman}(1984)}]{coleman1984new}
\bibinfo{author}{\bibfnamefont{P.}~\bibnamefont{Coleman}},
  \bibinfo{journal}{Phys. Rev. B} \textbf{\bibinfo{volume}{29}},
  \bibinfo{pages}{3035} (\bibinfo{year}{1984}).

\bibitem[{\citenamefont{Kotliar and Ruckenstein}(1986)}]{kotliar1986new}
\bibinfo{author}{\bibfnamefont{G.}~\bibnamefont{Kotliar}} \bibnamefont{and}
  \bibinfo{author}{\bibfnamefont{A.}~\bibnamefont{Ruckenstein}},
  \bibinfo{journal}{Phys. Rev. Lett.} \textbf{\bibinfo{volume}{57}},
  \bibinfo{pages}{1362} (\bibinfo{year}{1986}).

\bibitem[{\citenamefont{Kotliar and Liu}(1988)}]{kotliar1988superexchange}
\bibinfo{author}{\bibfnamefont{G.}~\bibnamefont{Kotliar}} \bibnamefont{and}
  \bibinfo{author}{\bibfnamefont{J.}~\bibnamefont{Liu}},
  \bibinfo{journal}{Phys. Rev. B} \textbf{\bibinfo{volume}{38}},
  \bibinfo{pages}{5142} (\bibinfo{year}{1988}).

\bibitem[{\citenamefont{Lee et~al.}(1998)\citenamefont{Lee, Nagaosa, Ng, and
  Wen}}]{lee19982}
\bibinfo{author}{\bibfnamefont{P.~A.} \bibnamefont{Lee}},
  \bibinfo{author}{\bibfnamefont{N.}~\bibnamefont{Nagaosa}},
  \bibinfo{author}{\bibfnamefont{T.-K.} \bibnamefont{Ng}}, \bibnamefont{and}
  \bibinfo{author}{\bibfnamefont{X.-G.} \bibnamefont{Wen}},
  \bibinfo{journal}{Phys. Rev. B} \textbf{\bibinfo{volume}{57}},
  \bibinfo{pages}{6003} (\bibinfo{year}{1998}).

\bibitem[{\citenamefont{Ruckenstein et~al.}(1987)\citenamefont{Ruckenstein,
  Hirschfeld, and Appel}}]{PhysRevB.36.857}
\bibinfo{author}{\bibfnamefont{A.~E.} \bibnamefont{Ruckenstein}},
  \bibinfo{author}{\bibfnamefont{P.~J.} \bibnamefont{Hirschfeld}},
  \bibnamefont{and} \bibinfo{author}{\bibfnamefont{J.}~\bibnamefont{Appel}},
  \bibinfo{journal}{Phys. Rev. B} \textbf{\bibinfo{volume}{36}},
  \bibinfo{pages}{857} (\bibinfo{year}{1987}).

\bibitem[{\citenamefont{Abrikosov et~al.}(1975)\citenamefont{Abrikosov, Gorkov,
  and Dzyaloshinski}}]{abrikosov1975methods}
\bibinfo{author}{\bibfnamefont{A.}~\bibnamefont{Abrikosov}},
  \bibinfo{author}{\bibfnamefont{L.}~\bibnamefont{Gorkov}}, \bibnamefont{and}
  \bibinfo{author}{\bibfnamefont{I.}~\bibnamefont{Dzyaloshinski}},
  \emph{\bibinfo{title}{Methods of quantum field theory in statistical
  physics}} (\bibinfo{publisher}{Courier Dover Publications},
  \bibinfo{year}{1975}).

\bibitem[{Sup()}]{Suppl}
\bibinfo{note}{See the Supplemental Material at
  http://link.aps.org/supplemental/XXXXX}.

\bibitem[{\citenamefont{Balatsky and Salkola}(1996)}]{PhysRevLett.76.2386}
\bibinfo{author}{\bibfnamefont{A.~V.} \bibnamefont{Balatsky}} \bibnamefont{and}
  \bibinfo{author}{\bibfnamefont{M.~I.} \bibnamefont{Salkola}},
  \bibinfo{journal}{Phys. Rev. Lett.} \textbf{\bibinfo{volume}{76}},
  \bibinfo{pages}{2386} (\bibinfo{year}{1996}).

\bibitem[{\citenamefont{Figgins and Morr}(2011)}]{FigginsMorr2011}
\bibinfo{author}{\bibfnamefont{J.}~\bibnamefont{Figgins}} \bibnamefont{and}
  \bibinfo{author}{\bibfnamefont{D.~K.} \bibnamefont{Morr}},
  \bibinfo{journal}{Phys. Rev. Lett.} \textbf{\bibinfo{volume}{107}},
  \bibinfo{pages}{066401} (\bibinfo{year}{2011}).

\end{thebibliography}


\begin{thebibliography}{1}
\expandafter\ifx\csname natexlab\endcsname\relax\def\natexlab#1{#1}\fi
\expandafter\ifx\csname bibnamefont\endcsname\relax
  \def\bibnamefont#1{#1}\fi
\expandafter\ifx\csname bibfnamefont\endcsname\relax
  \def\bibfnamefont#1{#1}\fi
\expandafter\ifx\csname citenamefont\endcsname\relax
  \def\citenamefont#1{#1}\fi
\expandafter\ifx\csname url\endcsname\relax
  \def\url#1{\texttt{#1}}\fi
\expandafter\ifx\csname urlprefix\endcsname\relax\def\urlprefix{URL }\fi
\providecommand{\bibinfo}[2]{#2}
\providecommand{\eprint}[2][]{\url{#2}}

\bibitem[{\citenamefont{Lee et~al.}(2006)\citenamefont{Lee, Nagaosa, and
  Wen}}]{lee2006doping}
\bibinfo{author}{\bibfnamefont{P.}~\bibnamefont{Lee}},
  \bibinfo{author}{\bibfnamefont{N.}~\bibnamefont{Nagaosa}}, \bibnamefont{and}
  \bibinfo{author}{\bibfnamefont{X.}~\bibnamefont{Wen}}, \bibinfo{journal}{Rev.
  Mod. Phys.} \textbf{\bibinfo{volume}{78}}, \bibinfo{pages}{17}
  (\bibinfo{year}{2006}).

\end{thebibliography}

\end{document}


\title{Mottness-induced healing in strongly correlated superconductors:
\\
supplemental material}

\author{Shao Tang }

\affiliation{Department of Physics and National High Magnetic Field Laboratory,
Florida State University, Tallahassee, Florida 32306, USA}

\author{E. Miranda}

\affiliation{Instituto de Física Gleb Wataghin, Campinas State University, Rua
Sérgio Buarque de Holanda, 777, CEP 13083-859, Campinas, Brazil}

\author{V. Dobrosavljevic}

\affiliation{Department of Physics and National High Magnetic Field Laboratory,
Florida State University, Tallahassee, Florida 32306, USA}

\pacs{71.10.Fd, 71.27.+a, 71.30.+h}

\maketitle

\section{The linear approximation}

\label{sec:linearappox}

Our linear approximation approach consists of expanding the mean-field
equations (2-5) of the main text to first order in the site energies
$\varepsilon_{i}$. Denoting linear deviations in the various fields
by $\delta$ we get

\begin{eqnarray}
\delta\chi_{ij} & = & 2kT\sum_{nl}\left(-g_{il}g_{lj}+G_{1il}G_{1lj}\right)\left(\delta\lambda_{l}+\varepsilon_{l}\right)+2kTr\sum_{nlm}\left(-g_{il}g_{mj}+G_{1il}G_{1mj}\right)\left(\delta r_{l}h_{lm}+\delta r_{m}h_{lm}\right)\nonumber \\
 &  & -2kT\sum_{nlm}\left(-g_{il}g_{mj}+G_{1il}G_{1mj}\right)\left(\widetilde{J}\delta\chi_{lm}\right)+2kT\sum_{nlm}\left(g_{il}G_{1mj}+g_{mj}G_{1il}\right)\left(\widetilde{J}\delta\Delta_{lm}\right),\label{eq:deltachi}
\end{eqnarray}

\begin{eqnarray}
\delta\Delta_{ij} & = & -2kT\sum_{nl}\left(g_{il}G_{1lj}+g_{lj}G_{1il}\right)\left(\delta\lambda_{l}+\varepsilon_{l}\right)-2kTr\sum_{nlm}\left(g_{il}G_{1mj}+g_{mj}G_{1il}\right)\left(\delta r_{l}h_{lm}+\delta r_{m}h_{lm}\right)\nonumber \\
 &  & +2kT\sum_{nlm}\left(g_{il}G_{1mj}+g_{mj}G_{1il}\right)\left(\widetilde{J}\delta\chi_{lm}\right)-2kT\sum_{nlm}\left(G_{1il}G_{2mj}+g_{il}g_{mj}\right)\left(\widetilde{J}\delta\Delta_{lm}\right),\label{eq:deltadelta}
\end{eqnarray}

\begin{eqnarray}
-r\delta r_{i} & = & kT\sum_{nl}\left(-g_{il}g_{li}+G_{1il}G_{1li}\right)\left(\delta\lambda_{l}+\varepsilon_{l}\right)+kTr\sum_{nlm}\left(-g_{il}g_{mi}+G_{1il}G_{1mi}\right)\left(\delta r_{l}h_{lm}+\delta r_{m}h_{lm}\right)\nonumber \\
 &  & -kT\sum_{nlm}\left(-g_{il}g_{mi}+G_{1il}G_{1mi}\right)\left(\widetilde{J}\delta\chi_{lm}\right)+kT\sum_{nlm}\left(g_{il}G_{1mi}+g_{mi}G_{1il}\right)\left(\widetilde{J}\delta\Delta_{lm}\right),\label{eq:deltar}
\end{eqnarray}

\begin{eqnarray}
\lambda\delta r_{i}+r\delta\lambda_{i}+\sum_{l}h_{il}\chi_{il}\delta r_{l}+r\sum_{l}h_{il}\delta\chi_{il} & = & 0,\label{eq:deltalambda}
\end{eqnarray}
where $G_{1ij}\equiv[G_{ij}]_{11}$, $G_{2ij}\equiv[G_{ij}]_{22}$,
$g_{ij}\equiv[G_{ij}]_{12}=[G_{ij}]_{21}$ are the Green's functions
of the clean system, $n$ is the fermionic Matsubara frequency index
and $\widetilde{J}=\frac{3}{8}J$. The latter choice is made, in the
presence of correlations, so that the multi-channel Hubbard-Stratonovich
transformation we used reproduces, at the saddle-point level, the
mean-field results \cite{lee2006doping}%
\footnote{The usual choice $\widetilde{J}=\frac{1}{4}J$ does not change the
analytical results and would give rise to hardly noticeable changes
in the numerical plots.%
}. In general, the clean Green's functions in $\mathbf{k}$-space are
given by 
\begin{equation}
G_{1}(\omega_{n},\mathbf{k})=\frac{i\omega_{n}+e(\mathbf{k})}{(i\omega_{n})^{2}-e^{2}(\mathbf{k})-\widetilde{J}^{2}\Delta^{2}(\mathbf{k})},\label{eq:g1}
\end{equation}

\begin{equation}
G_{2}(\omega_{n},\mathbf{k})=\frac{i\omega_{n}-e(\mathbf{k})}{(i\omega_{n})^{2}-e^{2}(\mathbf{k})-\widetilde{J}^{2}\Delta^{2}(\mathbf{k})},\label{eq:g2}
\end{equation}

\begin{equation}
g(\omega_{n},\mathbf{k})=\frac{\widetilde{J}\Delta(\mathbf{k})}{(i\omega_{n})^{2}-e^{2}(\mathbf{k})-\widetilde{J}^{2}\Delta^{2}(\mathbf{k})},\label{eq:gsmall}
\end{equation}
where the renormalized dispersion is

\begin{eqnarray}
e(\mathbf{k}) & = & -2\left(xt+\chi\widetilde{J}\right)\left[\cos\left(k_{x}a\right)+\cos\left(k_{y}a\right)\right]-4xt^{\prime}\cos\left(k_{x}a\right)\cos\left(k_{y}a\right)-\mu,\label{eq:disp}
\end{eqnarray}
we have absorbed the clean $\lambda$ in the chemical potential, and
\begin{equation}
\Delta\left(\mathbf{k}\right)=2\Delta_{0}\left[\cos\left(k_{x}a\right)-\cos\left(k_{y}a\right)\right].\label{eq:gap}
\end{equation}
Notice that the dimensionful gap function is $\Delta_{phys}\left(\mathbf{k}\right)=\widetilde{J}\Delta\left(\mathbf{k}\right)$.
As we focus on the asymptotic long-range behavior of the different
fields, their variations are dominated by the corresponding clean-limit
symmetry channel. We therefore define local order parameters as $\delta\chi_{i}\equiv\frac{1}{2d}\sum_{j}\delta\chi_{ij}\Gamma(s)_{ij}$,
$\delta\Delta_{i}\equiv\frac{1}{2d}\sum_{j}\delta\Delta_{ij}\Gamma(d_{x^{2}-y^{2}})_{ij}$.
Thus, \emph{defining vectors and matrices in the lattice site basis
with bold-face letters}, Eqs.~(\ref{eq:deltachi}-\ref{eq:deltalambda})
can be recast as
\begin{equation}
\left(\boldsymbol{A}+r^{2}\boldsymbol{B}\right)\delta\boldsymbol{\Phi}=r^{2}\boldsymbol{C},\label{eq:matrixequations}
\end{equation}
where 
\begin{equation}
\boldsymbol{A}=\left(\begin{array}{cccc}
\boldsymbol{M}_{11} & \boldsymbol{M}_{12} & \boldsymbol{M}_{13} & \boldsymbol{M}_{14}\\
\boldsymbol{M}_{21} & \boldsymbol{M}_{22} & \boldsymbol{M}_{23} & \boldsymbol{M}_{24}\\
\boldsymbol{M}_{31} & \boldsymbol{M}_{32} & \boldsymbol{M}_{33} & \boldsymbol{M}_{34}\\
0 & 0 & \lambda\mathbf{1}-\frac{\lambda}{2d}\boldsymbol{\Gamma}(s) & 0
\end{array}\right),\boldsymbol{B}=\left(\begin{array}{cccc}
0 & 0 & 0 & 0\\
0 & 0 & 0 & 0\\
0 & 0 & 0 & 0\\
-2dt\mathbf{1} & 0 & 0 & \mathbf{1}
\end{array}\right),\delta\boldsymbol{\Phi}=\left(\begin{array}{c}
\delta\boldsymbol{\chi}\\
\delta\boldsymbol{\Delta}\\
r\delta\boldsymbol{r}\\
\delta\boldsymbol{\overline{\lambda}}
\end{array}\right),\boldsymbol{C}=\left(\begin{array}{c}
0\\
0\\
0\\
\boldsymbol{\varepsilon}
\end{array}\right).\label{eq:matrices}
\end{equation}
Here, the elements of (the vector) $\boldsymbol{\varepsilon}$ are
the disorder potential values $\varepsilon_{i}$, $\mathbf{1}$ is
the identity matrix, $\delta\overline{\lambda}_{i}=\delta\lambda_{i}+\varepsilon_{i}$,
and

\begin{equation}
M_{11ij}=-\delta_{ij}-\frac{\widetilde{J}kT}{d}\sum_{nml}\Gamma(s)_{il}\left(-g_{ij}g_{ml}+G_{1ij}G_{1ml}\right)\Gamma(s)_{jm}\label{eq:m11}
\end{equation}

\begin{eqnarray}
M_{12ij} & = & \frac{\widetilde{J}kT}{d}\sum_{nml}\Gamma(s)_{il}\left(g_{ij}G_{1ml}+g_{ml}G_{1ij}\right)\Gamma(d_{x^{2}-y^{2}})_{jm}
\end{eqnarray}

\begin{eqnarray}
M_{13ij} & = & \frac{kT}{d}\sum_{nml}\Gamma(s)_{il}\left(-g_{ij}g_{ml}+G_{1ij}G_{1ml}-g_{im}g_{jl}+G_{1im}G_{1jl}\right)h_{jm}
\end{eqnarray}

\begin{equation}
M_{14ij}=\frac{kT}{d}\sum_{nl}\Gamma(s)_{il}\left(-g_{ij}g_{jl}+G_{1ij}G_{1jl}\right)
\end{equation}

\begin{eqnarray}
M_{21ij} & = & -\frac{\widetilde{J}kT}{d}\sum_{nml}\Gamma(d_{x^{2}-y^{2}})_{il}\left(g_{ij}G_{1ml}+g_{ml}G_{1ij}\right)\Gamma(s)_{jm}
\end{eqnarray}

\begin{eqnarray}
M_{22ij} & =\delta_{ij}+ & \frac{\widetilde{J}kT}{d}\sum_{nml}\Gamma(d_{x^{2}-y^{2}})_{il}\left(G_{1ij}G_{2ml}+g_{ij}g_{ml}\right)\Gamma(d_{x^{2}-y^{2}})_{jm}
\end{eqnarray}

\begin{eqnarray}
M_{23ij} & = & \frac{kT}{d}\sum_{nml}\Gamma(d_{x^{2}-y^{2}})_{il}\left(g_{ij}G_{1ml}+g_{ml}G_{1ij}+g_{im}G_{1jl}+g_{jl}G_{1im}\right)h_{jm}
\end{eqnarray}

\begin{equation}
M_{24ij}=\frac{kT}{d}\sum_{nl}\Gamma(d_{x^{2}-y^{2}})_{il}\left(g_{ij}G_{1jl}+g_{jl}G_{1ij}\right)
\end{equation}

\begin{eqnarray}
M_{31ij} & = & -\widetilde{J}kT\sum_{nm}\left(-g_{ij}g_{mi}+G_{1ij}G_{1mi}\right)\Gamma(s)_{jm}
\end{eqnarray}

\begin{equation}
M_{32ij}=\widetilde{J}kT\sum_{nm}\left(g_{ij}G_{1mi}+g_{mi}G_{1ij}\right)\Gamma(d_{x^{2}-y^{2}})_{jm}
\end{equation}

\begin{eqnarray}
M_{33ij} & = & \delta_{ij}+kT\sum_{nm}h_{jm}\left(-g_{ij}g_{mi}+G_{1ij}G_{1mi}-g_{im}g_{ji}+G_{1im}G_{1ji}\right)
\end{eqnarray}

\begin{equation}
M_{34ij}=kT\sum_{n}\left(-g_{ij}g_{ji}+G_{1ij}G_{1ji}\right)\label{eq:m34}
\end{equation}

In writing Eqs.~(\ref{eq:matrixequations}), we have made explicit
the $r$ dependence of Eqs.~(\ref{eq:deltachi}-\ref{eq:deltalambda}).
We note, however, that there is also an implicit dependence on $r$
through the dispersion (\ref{eq:disp}) (where $x=r^{2}$), which
enters the various Green's functions in Eqs.~(\ref{eq:g1}-\ref{eq:gsmall}).

Since the matrix elements in Eqs.~(\ref{eq:m11}-\ref{eq:m34}) are
all calculated in the \emph{translation-invariant clean system}, Eqs.~(\ref{eq:matrixequations})
can be easily solved in $\mathbf{k}$-space by matrix inversion. Normal
state results are obtained by removing the second row and column and
setting $\Delta\left(\mathbf{k}\right)$ to zero. Non-correlated results
correspond to the absence of slave bosons and constraints, so we just
remove the third and fourth rows and columns and set $x=1$ and $\lambda_{i}=0$.
In every case, the clean limit is first solved self-consistently for
$\chi$, $\Delta$, $\lambda$ and $\mu$, and then the fluctuations
in the presence of impurities are obtained.

In discussing the solution to Eqs.~(\ref{eq:matrixequations}), we
rely on the fact that all quantities in Eqs.~(\ref{eq:m11}-\ref{eq:m34})
are non-singular and finite as $x\to0$. Thus, we can write their
formal solution as
\begin{equation}
\delta\boldsymbol{\Phi}=r^{2}\left(\boldsymbol{A}+r^{2}\boldsymbol{B}\right)^{-1}\boldsymbol{C}=r^{2}\boldsymbol{A}^{-1}\boldsymbol{C}+{\cal O}\left(r^{4}\right).\label{eq:solution}
\end{equation}
\emph{It follows that $\delta\chi_{i}$, $\delta\Delta_{i}$, $r\delta r_{i}$,
and $\delta\overline{\lambda}_{i}=\delta\lambda_{i}+\varepsilon_{i}$
are all of order $r^{2}=x$.}

\section{The gap fluctuations and the healing factor}

\label{sec:gapfluc}

\begin{figure}
\includegraphics[scale=0.7]{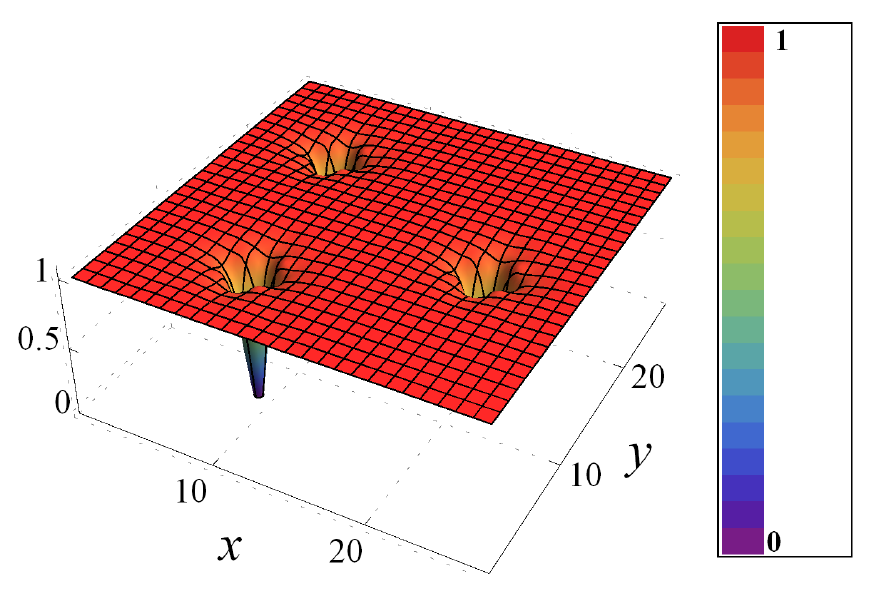}\includegraphics[scale=1.3]{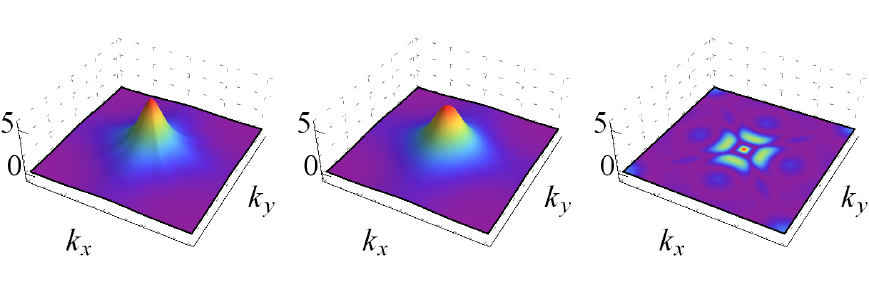}

\includegraphics[scale=0.7]{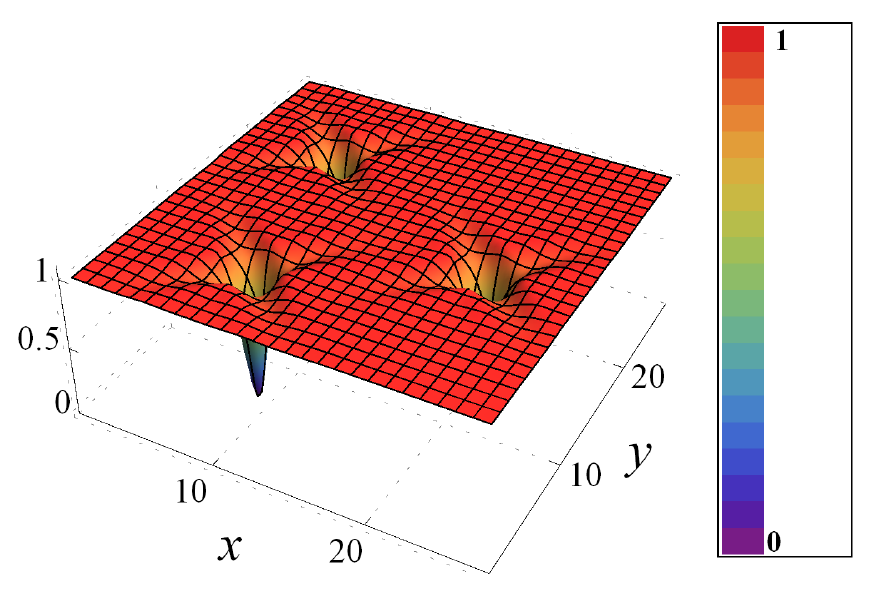}\includegraphics[scale=1.3]{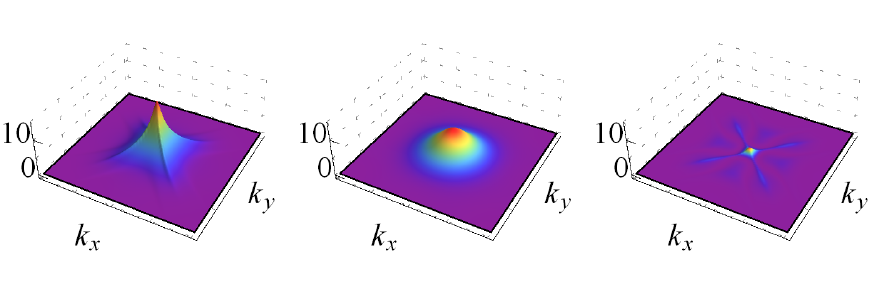}

\includegraphics[scale=0.7]{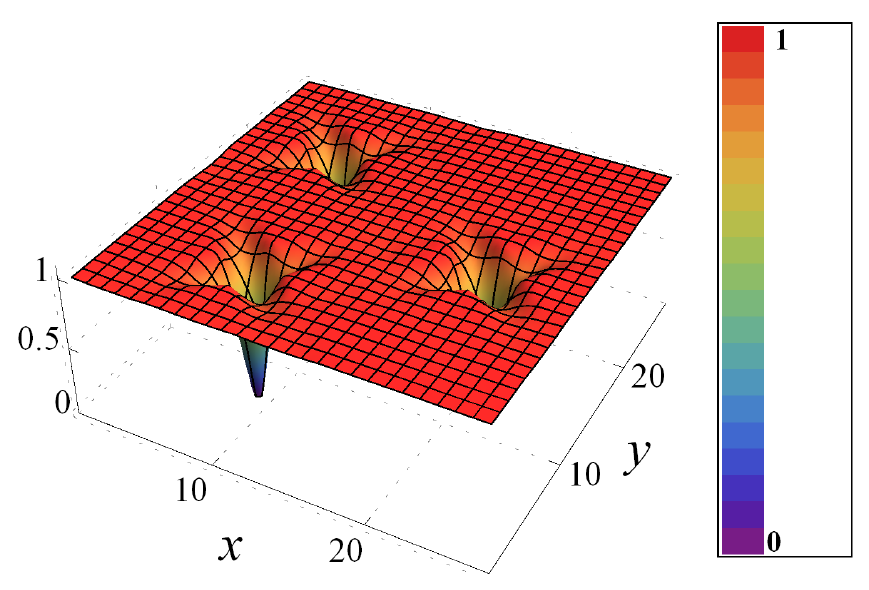}\includegraphics[scale=1.3]{0\lyxdot 3_powerspectrum}

\protect\caption{\label{fig:gapfluc=000026PSwithcorr}Spatial variations of normalized
local gap function $\frac{\delta\Delta_{i}}{\Delta_{0}}$ for three
impurities (first column) and the corresponding power spectra $S(\mathbf{k})$,
$S(\mathbf{k})_{loc}$ and $S(\mathbf{k})_{nonloc}$ (second to fourth
columns) for $x=0.15$ (first row), $x=0.25$ (second row), and $x=0.3$
(third row). The corresponding healing factors are (a) $h=0.23\%$,
(b) $h=1.77\%$, and (c) $h=2.74\%$.}
\end{figure}

In order to characterize quantitatively the healing process in the
SC state, we focused on the linear gap response to the disorder potential
\begin{equation}
\delta\Delta_{i}=\Delta_{0}\sum_{j}M_{\Delta}\left(\mathbf{r}_{i}-\mathbf{r}_{j}\right)\varepsilon_{j},
\end{equation}
which is obtained directly from the second line of the solution to
Eqs.~(\ref{eq:solution}). In order to gain further insight, we separated
the local and non-local parts of the gap response as follows. In Eqs.~(\ref{eq:matrixequations})
as defined in real space, we separate sums over sites into a local
part, with sums up to next-to-nearest neighbors (denoted by $r_{ij}\le\sqrt{2}a$),
and a non-local part, with sums over the remaining sites (denoted
by $r_{ij}>\sqrt{2}a$). For example,
\begin{equation}
\sum_{j}M_{11ij}\delta\chi_{j}=\sum_{j,r_{ij}\le\sqrt{2}a}M_{11ij}\delta\chi_{j}+\sum_{j,r_{ij}>\sqrt{2}a}M_{11ij}\delta\chi_{j},\,\mathrm{etc.}
\end{equation}
After solving the equations, this separation naturally defines local
and non-local responses of the various fields. In $\mathbf{k}$-space,
we can write
\begin{equation}
M_{\Delta}\left(\mathbf{k}\right)=M_{\Delta,loc}\left(\mathbf{k}\right)+M_{\Delta,nonloc}\left(\mathbf{k}\right).
\end{equation}
This procedure is equivalent to projecting the full response in $\mathbf{k}$-space
onto some lattice symmetry channels with different ranges: $\Gamma_{s}\left(\mathbf{k}\right)=2\left[\cos\left(k_{x}a\right)+\cos\left(k_{y}a\right)\right]$
for nearest neighbors, and so on. Then, the power spectrum of spatial
gap fluctuations follows naturally from this separation
\begin{eqnarray}
S\left(\mathbf{k}\right) & = & M_{\Delta}^{2}\left(\mathbf{k}\right),\\
S_{loc}\left(\mathbf{k}\right) & = & M_{\Delta,loc}^{2}\left(\mathbf{k}\right),\\
S_{nonloc}\left(\mathbf{k}\right) & = & \left[M_{\Delta}\left(\mathbf{k}\right)-M_{\Delta,loc}\left(\mathbf{k}\right)\right]^{2}.
\end{eqnarray}
Finally, we define the healing factor as the ratio of integrated non-local
to local contributions to the power spectrum
\begin{equation}
h=\frac{\int S_{nonloc}\left(\mathbf{k}\right)d^{2}k}{\int S_{loc}\left(\mathbf{k}\right)d^{2}k}.
\end{equation}
The gap fluctuations $\delta\Delta_{i}$ for three impurities and
power spectra, for several dopings and in the presence of correlations,
are shown in Fig.~\ref{fig:gapfluc=000026PSwithcorr}. The strong
healing in the presence of correlations is conspicuous. It is important
to note that this suppression of gap fluctuations is not restricted
to small dopings and remains quite strong even at $x=0.3$, where
the healing factor does not exceed 3\%. As explained in the main text,
the healing effect originates in the dominance of the local spherically
symmetric part (third column in Fig.~\ref{fig:gapfluc=000026PSwithcorr})
over the anisotropic non-local response (fourth column in Fig.~\ref{fig:gapfluc=000026PSwithcorr}).

\section{The irrelevance of spinon fluctuations and the ``minimal model''}

\label{sec:irrelevancespinon}

We can shed light on the strong healing effect by studying a simplified
case obtained by ``turning off'' the $\delta\chi_{i}$ fluctuations.
In this case, we need to solve the smaller set of equations
\begin{equation}
\left(\begin{array}{ccc}
\boldsymbol{M}_{22} & \boldsymbol{M}_{23} & \boldsymbol{M}_{24}\\
\boldsymbol{M}_{32} & \boldsymbol{M}_{33} & \boldsymbol{M}_{34}\\
0 & \lambda\mathbf{1}-\frac{\lambda}{2d}\boldsymbol{\Gamma}(s) & r^{2}\mathbf{1}
\end{array}\right)\left(\begin{array}{c}
\delta\boldsymbol{\Delta}\\
r\delta\boldsymbol{r}\\
\delta\overline{\boldsymbol{\lambda}}
\end{array}\right)=\left(\begin{array}{c}
0\\
0\\
r^{2}\boldsymbol{\varepsilon}
\end{array}\right).\label{eq:SCnochi}
\end{equation}
The healing factor obtained in this simplified model is almost identical
to the full solution, as shown by the red and green curves of the
left panel of Fig.~2 of the main text. This shows that the spinon
field fluctuations are utterly irrelevant for the strong healing.

A further fruitful simplification is obtained by setting $M_{32}$
to zero in Eqs.~(\ref{eq:SCnochi}). This defines what we called
the ``minimal model'' (MM). In this case, the ``strong-correlation
sub-block'' of $\delta r_{i}$ and $\delta\overline{\lambda}_{i}$
fluctuations decouples and suffers no feed-back from the gap fluctuations.
In fact, the MM corresponds to breaking up the solution to the problem
into two parts: (i) the spatially fluctuating strong correlation fields
$r_{i}$ and $\lambda_{i}$ are first calculated for \emph{fixed,
uniform} $\Delta$ and $\chi$, and then (ii) the effects of their
spatial readjustments are fed back into the gap equation to find $\delta\Delta_{i}$. 

\begin{figure}
\includegraphics[scale=1.3]{0\lyxdot 2_powerspectrum}

\includegraphics[scale=1.3]{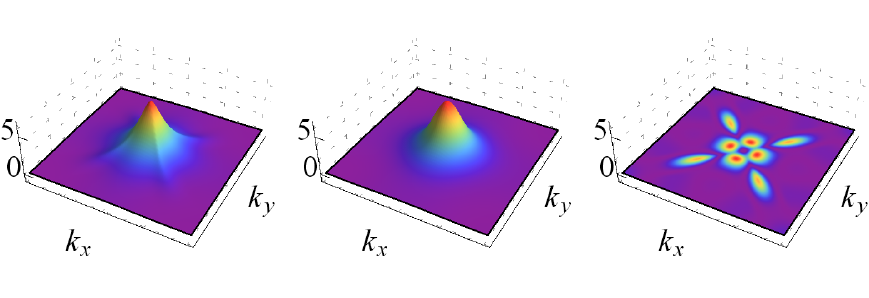}\protect\caption{\label{fig:gapflucMM}Power spectra of gap fluctuations $S(\mathbf{k})$,
$S(\mathbf{k})_{loc}$ and $S(\mathbf{k})_{nonloc}$ (first to third
columns) for $x=0.2$ in the presence of correlations. The top figures
were obtained from the full solution of the linearized Eqs.~(\ref{eq:matrixequations}),
whereas the bottom ones correspond to the minimal model (Eqs.~(\ref{eq:gapflucminimal})).
Note that the healing factors are $h=0.74\%$ (top) and $h=0.69\%$
(bottom).}
\end{figure}

Strikingly, the healing factor in this case is \emph{numerically indistinguishable}
from the one obtained from Eqs.~(\ref{eq:SCnochi}) (green curve
of the left panel of Fig.~2 of the main text). Furthermore, the full,
local and non-local PS of gap fluctuations are also captured quite
accurately by the MM, as seen in Fig.~\ref{fig:gapflucMM}. We conclude
that the MM, which incorporates only the effects of strong correlations,
is able to describe with very high accuracy the healing process in
the $d$-wave SC state.

The MM also permits us to obtain simple and physically transparent
expressions. In particular, it follows immediately that 
\begin{equation}
r\delta r\left(\mathbf{k}\right)=\frac{r^{2}}{\lambda a\left(\mathbf{k}\right)-r^{2}M_{33}\left(\mathbf{k}\right)/M_{34}\left(\mathbf{k}\right)}\varepsilon\left(\mathbf{k}\right),\label{eq:rdeltar}
\end{equation}
where $a\left(\mathbf{k}\right)=1-\Gamma_{s}\left(\mathbf{k}\right)/4$,
and we used the Fourier transform of $\boldsymbol{\Gamma}(s)$, $\Gamma_{s}\left(\mathbf{k}\right)=2\left[\cos\left(k_{x}a\right)+\cos\left(k_{y}a\right)\right]$.
Moreover, 
\begin{eqnarray}
\delta\Delta\left(\mathbf{k}\right) & = & \frac{r^{2}\left[M_{24}\left(\mathbf{k}\right)M_{33}\left(\mathbf{k}\right)-M_{23}\left(\mathbf{k}\right)M_{34}\left(\mathbf{k}\right)\right]}{M_{22}\left(\mathbf{k}\right)\left[\lambda a\left(\mathbf{k}\right)M_{34}\left(\mathbf{k}\right)-r^{2}M_{33}\left(\mathbf{k}\right)\right]}\varepsilon\left(\mathbf{k}\right),\label{eq:gapflucminimal}\\
 & = & \frac{\left[M_{24}\left(\mathbf{k}\right)\frac{M_{33}\left(\mathbf{k}\right)}{M_{34}\left(\mathbf{k}\right)}-M_{23}\left(\mathbf{k}\right)\right]}{M_{22}\left(\mathbf{k}\right)}r\delta r\left(\mathbf{k}\right).\label{eq:gapflucminimal2}\\
 & = & \chi_{pc}^{MM}\left(\mathbf{k}\right)\delta n\left(\mathbf{k}\right),\label{eq:gapflucminimal3}
\end{eqnarray}
where we used $n_{i}=1-r_{i}^{2}\Rightarrow\delta n_{i}=-2r\delta r_{i}$,
and 
\begin{equation}
\chi_{pc}^{MM}\left(\mathbf{k}\right)=-\frac{\left[M_{24}\left(\mathbf{k}\right)\frac{M_{33}\left(\mathbf{k}\right)}{M_{34}\left(\mathbf{k}\right)}-M_{23}\left(\mathbf{k}\right)\right]}{2M_{22}\left(\mathbf{k}\right)}.\label{eq:chipcMM}
\end{equation}
The local part of the response, which we have shown to be the dominant
one, can be studied by looking at the long wavelength limit. As $k\to0$,
$a\left(\mathbf{k}\right)\sim k^{2}/4$ and 
\begin{equation}
\delta\Delta_{loc}\left(\mathbf{k}\right)\approx-\chi_{pc}^{MM}\left(\mathbf{k}=0\right)\frac{8r^{2}/\lambda}{k^{2}+\xi_{S}^{-2}}\varepsilon\left(\mathbf{k}\right),\label{eq:gapfluclorentzian}
\end{equation}
where
\begin{equation}
\frac{1}{\xi_{S}}=\sqrt{-\frac{4r^{2}}{\lambda}\frac{M_{33}\left(\mathbf{k}=0\right)}{M_{34}\left(\mathbf{k}=0\right)}}.\label{eq:csiSMM}
\end{equation}
Eqs.~(\ref{eq:chipcMM}) and (\ref{eq:csiSMM}) give us the expressions
for the pair-charge correlation function and the healing length within
the MM.

\section{The normal state and the ``minimal model''}

\label{sec:normalstate}

\begin{figure*}[t]
\includegraphics[scale=0.7]{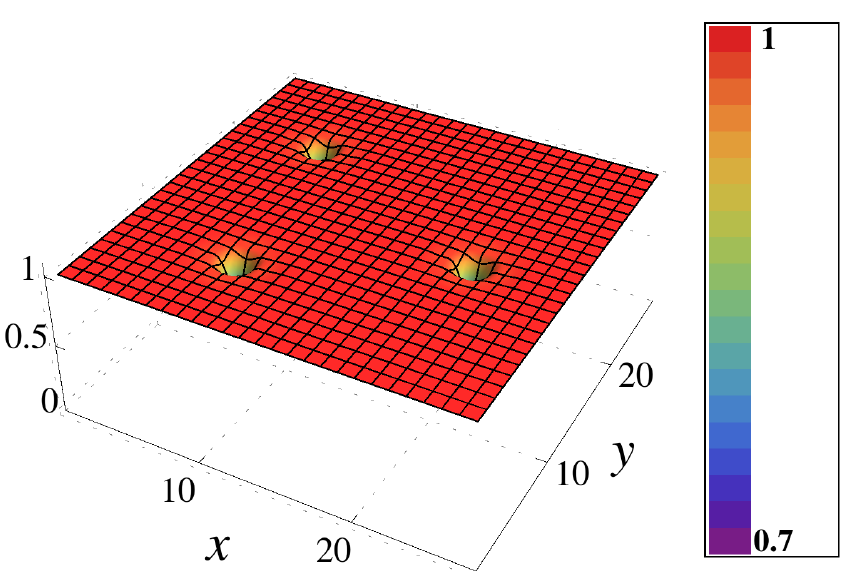}\includegraphics[scale=1.3]{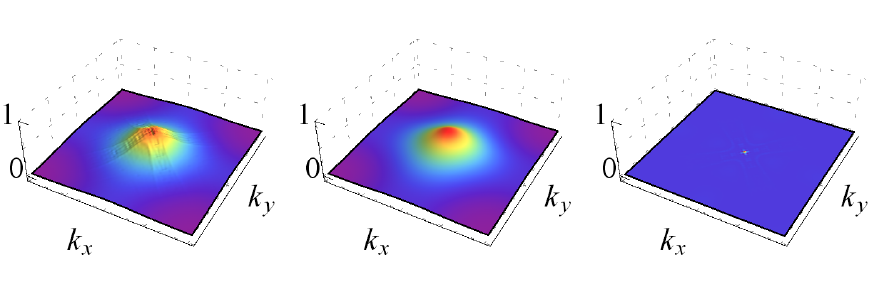}

\includegraphics[scale=0.7]{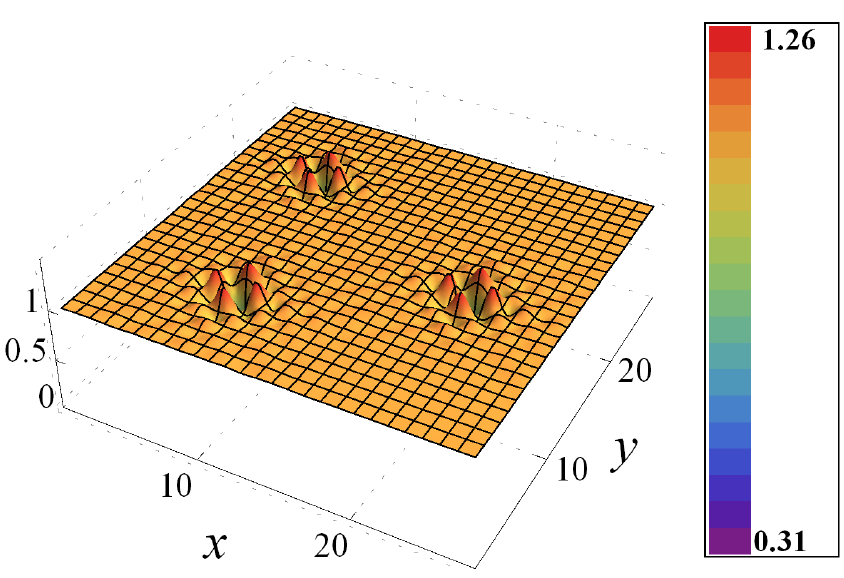}\includegraphics[scale=1.3]{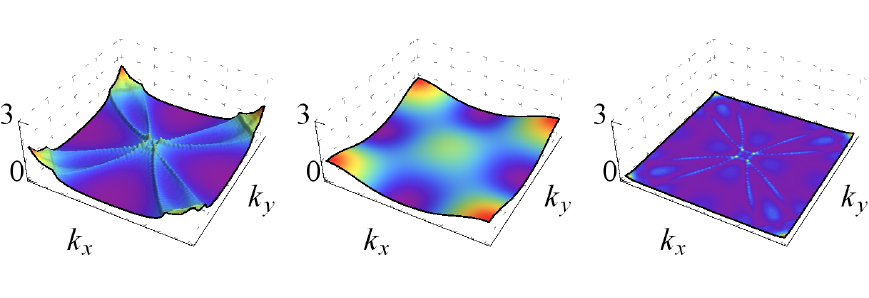}\protect\caption{\label{fig:densityfluct}Spatial variations of normalized local density
$\frac{\delta n_{i}}{n_{0}}$ in the normal state for three impurities
(first column) and the corresponding power spectra $N(\mathbf{k})$,
$N(\mathbf{k})_{loc}$ and $N(\mathbf{k})_{nonloc}$ (second to fourth
columns), in the presence (top) and in the absence (bottom) of strong
correlations for $x=0.2$. The strong suppression of density oscillations
by correlations is accompanied by the dominance of the spherically
symmetric local power spectrum {[}$N_{loc}\left(\mathbf{k}\right)${]}
over the anisotropic non-local one {[}$N_{nonloc}\left(\mathbf{k}\right)${]}.}
\end{figure*}

It is instructive to analyze also the behavior of the charge fluctuations
in the normal state. This can be achieved by suppressing the second
row and column of Eqs.~(\ref{eq:matrixequations}) and setting $\Delta\left(\mathbf{k}\right)$,
and thus $g\left(i\omega_{n},\mathbf{k}\right)$, to zero. Even after
these simplifications, the full solution is long and cumbersome. However,
accurate insight can be gained from a MM of the normal state, in which
we also set the $\delta\chi_{i}$ to zero by hand. As before, the
strong-correlation sub-block decouples and Eq.~(\ref{eq:rdeltar})
is still valid (albeit with matrix elements calculated in the normal
state). The local part of the charge response is given by an expression
similar to Eq.~(\ref{eq:gapfluclorentzian})
\begin{equation}
\delta n_{loc}\left(\mathbf{k}\right)\approx-\frac{8r^{2}/\lambda}{k^{2}+\xi_{N}^{-2}}\varepsilon\left(\mathbf{k}\right),\label{eq:chargefluclorentzian}
\end{equation}
where $\xi_{N}$ is given by Eq.~(\ref{eq:csiSMM}), again with matrix
elements calculated in the normal state. The behavior of $\xi_{N}$
as a function of doping is shown by the green curve of the right panel
of Fig.~2 of the main text.

In addition, just like in the Coulomb gas, the density fluctuations
also show Friedel-like oscillations coming from the singularity in
the response function at $2k_{F}$. Thus, expanding Eq.~(\ref{eq:rdeltar})
in $r^{2}$, 
\begin{equation}
\delta n_{nonloc}\left(\left|\mathbf{k}\right|\approx2k_{F}\right)\approx-\frac{2r^{2}}{\lambda a\left(\left|\mathbf{k}\right|\approx2k_{F}\right)}\left[1+\frac{r^{2}M_{33}\left(\left|\mathbf{k}\right|\approx2k_{F}\right)/M_{34}\left(\left|\mathbf{k}\right|\approx2k_{F}\right)}{\lambda a\left(\left|\mathbf{k}\right|\approx2k_{F}\right)}\right]\varepsilon\left(\mathbf{k}\right).\label{eq:chargeflucnonlocal}
\end{equation}
Since 
\begin{eqnarray}
M_{34}\left(\mathbf{k}\right) & = & \mathbf{\Pi}\left(\mathbf{k}\right),\\
M_{33}\left(\mathbf{k}\right) & = & 1+\mathbf{\Pi}^{b}\left(\mathbf{k}\right),\\
\mathbf{\Pi}\left(\mathbf{k}\right) & = & \frac{1}{V}\sum_{\mathbf{q}}\frac{f\left[\widetilde{h}\left(\mathbf{q}+\mathbf{k}\right)\right]-f\left[\widetilde{h}\left(\mathbf{q}\right)\right]}{\widetilde{h}\left(\mathbf{q}+\mathbf{k}\right)-\widetilde{h}\left(\mathbf{q}\right)},\\
\mathbf{\Pi}^{b}\left(\mathbf{k}\right) & = & \frac{1}{V}\sum_{\mathbf{q}}\frac{f\left[\widetilde{h}\left(\mathbf{q}+\mathbf{k}\right)\right]-f\left[\widetilde{h}\left(\mathbf{q}\right)\right]}{\widetilde{h}\left(\mathbf{q}+\mathbf{k}\right)-\widetilde{h}\left(\mathbf{q}\right)}\left[h\left(\mathbf{q}+\mathbf{k}\right)+h\left(\mathbf{q}\right)\right],\\
h\left(\mathbf{k}\right) & = & -t\Gamma_{s}\left(\mathbf{k}\right)-4t^{\prime}\cos\left(k_{x}a\right)\cos\left(k_{y}a\right),
\end{eqnarray}
the leading divergent behavior is 
\begin{equation}
\frac{M_{33}\left(\left|\mathbf{k}\right|\approx2k_{F}\right)}{M_{34}\left(\left|\mathbf{k}\right|\approx2k_{F}\right)}\approx\frac{1}{\mathbf{\Pi}\left(\left|\mathbf{k}\right|\approx2k_{F}\right)}.
\end{equation}
The two contributions from Eqs.~(\ref{eq:chargefluclorentzian})
and (\ref{eq:chargeflucnonlocal}) together give, in real space, 
\begin{eqnarray}
\frac{\delta n_{i}}{n_{0}} & = & x\sum_{j}\left(c_{1}\frac{e^{-r_{ij}/\xi}}{\xi^{(d-3)/2}(r_{ij})^{(d-1)/2}}+c_{2}x\left[\boldsymbol{\Pi^{-1}}\right]_{ij}\right)\varepsilon_{j},\label{eq:chargepert}
\end{eqnarray}
where $r_{ij}$ is the distance between sites $i$ and $j$, and $c_{1}$
and $c_{2}$ are constants that depend on $t,t^{\prime},J$ and $x$.

We stress that in the full solution of the linearized equations in
which $\delta\chi_{i}\neq0$, the structure of Eq.~(\ref{eq:rdeltar})
is still preserved, with the factor $M_{33}/M_{34}$ being replaced
by a long combination of several $M_{ij}$ elements, which, however,
has a \emph{finite negative }$\mathbf{k}\to0$ \emph{limit and a singularity
at} $2k_{F}$. Therefore, the results of Eqs.~(\ref{eq:chargefluclorentzian}),
(\ref{eq:chargeflucnonlocal}) and (\ref{eq:chargepert}) remain valid
in the general case. The spatial charge fluctuations for three impurities
and the PS in the normal state in the full solution are shown in Fig.~\ref{fig:densityfluct}
both in the absence and in the presence of strong correlations. Note
how the non-local part is down by an additional factor of $x$ as
compared to the local part {[}see Eqs.~(\ref{eq:chargeflucnonlocal})
and (\ref{eq:chargepert}){]}.

\bibliographystyle{apsrev}
\bibliography{manuscript}